\documentclass[pra,showpacs,floatfix,amsmath,amsfonts,twocolumn]{revtex4}
\usepackage{graphics}
\usepackage{epsfig}
\usepackage{amssymb}
\usepackage{color}

\begin{document}

\title{Tunable spin-orbit coupled Bose-Einstein condensates in deep optical lattices}

\author{ M. Salerno$^{1}$, F.Kh. Abdullaev$^{2,3}$, A. Gammal$^{4}$, Lauro Tomio$^{2,5,6}$ }
\affiliation{$^{1}$Dipartimento di Fisica ``E.R. Caianiello'', CNISM and INFN- Gruppo Collegato di Salerno, Universit\'a di Salerno, Via Giovanni
Paolo II, 84084 Fisciano (SA), Italy}
\affiliation{$^{2}$Centro de Ci\^encias Naturais e Humanas, Universidade Federal do ABC, 09210-170, Santo Andr\'e, Brazil}
\affiliation{$^{3}$Department of Physics, Kulliyyah of Science, International Islamic University Malaysia, 25200 Kuantan, Malaysia}
\affiliation{$^{4}$Instituto de F\'isica, Universidade de S\~{a}o Paulo, 05508-090 S\~{a}o Paulo, Brazil}
\affiliation{$^{5}$Instituto Tecnol\'ogico de Aero\'autica, CTA, 12228-900, S\~ao Jos\'e dos Campos, Brazil}
\affiliation{$^{6}$Instituto de F\'isica Te\'orica, Universidade Estadual Paulista (UNESP), 01140-070 S\~{a}o Paulo, Brazil}
\date{\today}
\begin{abstract}
Binary mixtures of Bose-Einstein condensates trapped in deep optical lattices and subjected to equal contributions of Rashba and
Dresselhaus spin-orbit coupling (SOC), are investigated  in the presence of a periodic time modulation of the Zeeman field.  SOC
tunability is explicitly demonstrated by adopting a mean-field  tight-binding model for the BEC mixture and by performing an averaging
approach in the strong modulation limit.
In this case, the system can be reduced to an unmodulated vector discrete nonlinear Schr\"odinger equation with a rescaled SOC
tunning parameter $\alpha$, which depends only on the ratio between amplitude and frequency of the applied Zeeman field.
The dependence of the spectrum of the linear system on $\alpha$ has been analytically characterized. In particular, we
show that extremal curves (ground and highest excited states) of the linear spectrum are continuous piecewise functions
(together with their derivatives) of $\alpha$, which consist of a finite number of decreasing band lobes joined by constant lines.
This structure also remains in presence of not too large nonlinearities. Most important, the interactions introduce a number of  localized states
in the band-gaps that  undergo change of properties as they collide with  band lobes. The stability of ground states in the presence of the
modulating field has been demonstrated by real time evolutions of the original (un-averaged) system. Localization properties of the ground
state induced by the SOC tuning,  and a parameter design for possible experimental observation have also been discussed.
\end{abstract}
\pacs{03.75.Lm, 03.75.Nt, 05.30.Jp}
\maketitle

\section{Introduction}
Spin-orbit coupling (SOC), i.e., the intrinsic interaction between the particle dynamics and its spin, is a phenomenon known from the dawn of quantum
mechanics, representing  a major source of magnetic  intra-atomic   interaction. In solid state physics, SOC plays an important role mainly in the
magnetism of solids that are well described in terms of individual  ions, as it is for the case for the earth-rare insulators, as well as in the study of energy
bands of semiconductors in the vicinity of the  extremal points where usually induces band splitting. The relevance of SOC in this context is well known
from pioneering works of Dresselhaus and Rashba~\cite{elliott,dresselhaus-kittel,dresselhaus,rashba,rashba2,Rashba-review} and from the many
theoretical and experimental developments which originated from them. In particular, in the recent past few decades there has been a flourishing of
interest in developments of materials with strong SOC for practical applications in the fields of  topological insulators~\cite{Hasan-Kane},
spintronics~\cite{Awschalom},
anomalous Hall effects~\cite{Nagaosa}, and quantum computation~\cite{stepanenko}, among other possibilities.

In  generic condensed matter  materials, however, SOC is rather weak and also very difficult to manage being largely superseded by the
electrostatic interactions. The situation is quite  different with ultra-cold atoms for which a variety of synthetic SOC can be induced and
managed by  external laser fields. In particular, SOC has been experimentally realized for binary mixtures
of Bose-Einstein condensates (BEC), as reported in Refs.~\cite{LJS,GS}, and theoretically investigated in several papers. The flexibility of
ultra-cold atomic systems in the control of  the interactions and the different types of  SOC implementations permit to explore   novel magnetic
phenomena difficult to  achieve with solid state materials. In this regard, we can mention the existence of  new  superfluid phases with unusual
magnetic properties~\cite{Lin-nature2009,zhu-epl}, stripe modes~\cite{stripe},
fractional topological insulators~\cite{phase,phase1,phase2}, new topological excitation such as  Weyl~\cite{weyl} and Majorana~\cite{maiorana}
fermions, antiferromagnetic states\cite{ZDKM13}, solitons\cite{sol1,sol2,sol3,sol4,KKZ} and gap solitons\cite{KKA-PRL13,KKZ14,YZhang15}.
In these contexts the tunability of SOC plays a crucial role both for distinguishing different phases arising under variations of parameters and for
understanding the mechanism underlying the phenomena as well as the interplay between SOC and the inter- and intra-atomic interactions (nonlinearity).

Recently a lot of attention has been devoted to the investigation of universal high-frequency behavior of periodically driven systems.
The   important consequences are dynamical stabilizations and the Floquet engineering of cold-atomic systems under temporal modulations
of parameters of the systems (in this regard, we can mention the recent review \cite{Bukov}).  One should also notice the number of 
theoretical ~\cite{Zhang} and experimental ~\cite{Spielman} studies on SOC tunability,
which have been done for continuous  BEC systems with equal Rashba and Dresselhaus terms, by using  rapid time variations
of the Raman frequency. In view of  the relevance of SOC induced phenomena, it  is interesting to explore SOC tunability also for different
parameter's modulations and  in the presence of   discrete settings as
the ones  induced by  the presence of  deep optical lattices (OL).

The aim of the  present paper is to investigate the SOC tunability of a binary BEC mixture trapped in a quasi one-dimensional deep optical
lattice in the presence of a time dependent Zeeman field. In this respect,  we consider SOC realized in the Weyl form
by means of  optical methods, using either the tripod scheme~\cite{tripod,EOUFTO} or four internal states with tetrahedral
geometry~\cite{Anderson}. The external Zeeman field is assumed to vary periodically in time, restricting  mainly to the
case in which the amplitude, $\Omega_1$, and frequency, $\omega$, of the modulation is very large (strongly modulation limit). The
deepness of the  OL is accounted by adopting the tight-binding SOC model of the BEC mixture introduced in Ref.~\cite{SA}, which is
in the form  of a vector discrete nonlinear Schr\"odinger equation (VDNLSE) with time dependent Zeeman field. We show that this model reduces
to an  effective time-averaged equation which has the same form as for the original unmodulated system, but with an effective SOC
parameter rescaled by a factor $J_0(\alpha)$, where $J_0$ is the zero-order Bessel function and $\alpha\equiv 2 \Omega_1/\omega$
is the tuning parameter.

The effect of the modulating field on the energy (chemical potential) spectrum is studied by exact analytical expressions in the absence of
 nonlinearity,  while we recourse to direct real and  imaginary  time evolutions of the original system and to exact self-consistent
numerical  diagonalization of the averaged Hamiltonian system in the nonlinear case.
In particular, we show that the ground state curve of the linear system is a piecewise function of $\alpha$ consisting equally-spaced branches (lobes)
centered around the relative minima of the chemical potential and joined by flat regions of constant $\mu$. A similar result applies also to  the  highest
excited extremal curve by symmetry arguments. In the presence of interactions, besides the removal of the degeneracy of extended states, a set of
discrete localized levels appear in the forbidden zone of the underlying linear band-gap structure, which displays oscillatory behaviors in terms of the
tuning parameter, with amplitudes that decrease as $\alpha$ is increasing The existence of ground-state stationary discrete solitons is explicitly
investigated both by means of exact diagonalizations of the averaged system and by direct imaginary time evolutions of the original system. The
stability of these states is  demonstrated by real time evolutions of the original (un-averaged) system. We also  consider the effect of the SOC tuning on
localization properties of the
ground state, by showing that for fixed equal attractive inter- and intra-species interactions there exists  an optimal value of $\alpha$ for which the
maximum localization of the wave function is achieved. This optimal tuning corresponds to the  point where the separation of the ground-state level
from the bottom of the linear band assumes its maximum value as a function of $\alpha$. The existence and stability of stripe-like soliton solutions are
also demonstrated. We find that, within the range of the $\alpha$ parameter and nonlinearity for which these solutions exist, their behaviors are similar
to the one obtained for stationary ground states. The possibility to observe these phenomena in real experiments is briefly discussed at the end.

The paper is organized as follows. In Sec. II, we introduce the  model equations of a  binary BEC mixture in a deep OL with SOC and modulating Zeeman
fields and derive the  averaged equations with rescaled SOC parameter. In Sec. III, we use the dispersion relation of the averaged linear system
to investigate the properties of the ground and highest excited states as functions of the tuning parameter. In  Sec. IV we study how the linear spectral
properties are affected by the nonlinearity. In Sec. V we study the influence of the SOC tuning on discrete soliton ground states with respect to existence
and stability, as well as localization properties. The stability of the results, under time integrations, are shown by considering full numerical simulation of
ground-state wave functions for different parameter choices. Finally, in Sec. VI, we discuss possible experimental implementations, physical estimates,
and conclude by summarizing our results.

\section{Model equations, averaging and SOC tuned linear spectrum}

The model equations for a  BEC mixture in a one-dimensional (1D) geometry can be derived from a more general three-dimensional
formalism by considering a trapping potential with the transversal frequency $\omega_\perp$ much larger than the longitudinal one,
$\omega_\perp\gg\omega_{||}$. In the present case, the trap potential in the $x-$direction is an optical lattice given by a periodic potential
$V_{ol}(x)\sim \sin^2(k_Lx)$, where $k_L$ is the lattice wave-number. In the mean field approximation, the system is described by a
1D Gross-Pitaevskii (GP) coupled equation for the two-component wave function, $\Psi\equiv\Psi(x,t)$, which is normalized to the total
number, $N$,  of atoms, as
\begin{equation}
\Psi\equiv \left( \begin{array}{c} \Psi_1\\ \Psi_2 \end{array} \right),\;\;\;
\sum_{j=1}^2 \int dx |\Psi_j|^2 =N
.\label{norm0}\end{equation}
In the presence of SOC the  corresponding GP formalism is given by the following one-dimensional (1D) Hamiltonian,
with two terms. The first term, $H_{0}$, is linear and includes the SOC and optical lattice.
The other term, given by $H_{nl}$, is non-linear and includes the two-body atomic interactions~\cite{ZMZ,ZDKM13,KKA-PRL13}.
In matricial form, it can be written as
{\small
\begin{eqnarray}\label{eqGP0}
i\hbar\frac{\partial\Psi}{\partial t} & =& \left[H_0+ H_{nl}\right]\Psi ,\nonumber \\
H_0&\equiv & \frac{P_x^2}{2m} +\frac{\hbar\kappa}{m} P_x\sigma_x + V_{ol}(x) +
\hbar {\overline\Omega}\sigma_z,\label{HNL}\\
H_{nl}&\equiv& 2\hbar\omega_\perp\left(
\begin{array}{cc}
\sum_j a_{1j}|\Psi_j|^2 & 0\\
0&\sum_ja_{j2}|\Psi_j|^2\\
\end{array}
\right)
 \nonumber
,\end{eqnarray}}
where $\sigma_{x,z}$ are the usual Pauli matrices, $a_{jj}\;\;(j=1,2)$ and $a_{12}$ are the two-body scattering lengths between
intra- and inter-species of atoms, and the parameter $\Omega_Z$ is defined by detuning or by the external Zeeman field.
The above formalism, with Eqs.~(\ref{norm0}) and (\ref{HNL}), can be written as
\begin{eqnarray}
i\hbar\frac{\partial \Psi_{j}}{\partial t} & =&\left[
-\frac{\hbar^2}{2m}\frac{\partial^2}{\partial x^2 }
+V_{ol}(x) -(-)^j\hbar{\overline\Omega}\right]\Psi_j
\nonumber\\
&+&2\hbar\omega_\perp(a_{jj}|\Psi_j|^2 + a_{j,3-j}|\Psi_{3-j}|^2)\Psi_j
\label{eqGP}\\
&-&i\frac{\hbar^2\kappa}{m}\frac{\partial\Psi_{3-j}}{\partial x}\;\;\;(j=0,1)
\nonumber .\end{eqnarray}
The form of SOC corresponding to this GP system can be obtained by using a tripod scheme~\cite{tripod,EOUFTO} for
the generation of synthetic gauge fields. The scheme operates with  atoms with three ground states $|i\rangle,i=1,2,3$
and one excited state $|e\rangle$, coupled by three laser beams $\Omega_{12}=(\Omega_0/\sqrt{2})\exp(-ik_2y\pm i\kappa x)\sin(\theta),
\Omega_3=\Omega_0 \exp(i\kappa z)$,  where $k_2$ and $\kappa$ are the wave vectors, with $\Omega_0$ and $\theta$
being amplitude and phase respectively.
The optical lattice, given by $V_{ol}(x)\equiv V_0 \cos(2k_L x)$,
can be generated by two counter-propagating laser fields.
To reach a dimensionless equation, we make the following replacements in  Eq.~(\ref{eqGP}):
\begin{eqnarray}
x&\to&\frac{x}{k_L},\;\; t\to \omega_R t,\;\;{\rm where}\;\; \omega_R\equiv \frac{E_R}{\hbar}\equiv \frac{\hbar k_L^2}{2m};\nonumber\\
 {V_{ol}(x)} &\to & E_R V(x) = E_R V_0\cos(2x),\label{ad-var} \\
 \Psi_j&\equiv&\sqrt{\frac{\omega_R}{2\omega_{\perp}a_0}}  \psi_j(x,t) ,\nonumber
 \end{eqnarray}
 with the definitions
 \begin{eqnarray}
b&\equiv& \frac{2\kappa}{k_L},\;\; \Omega_1=-\Omega_2=\frac{\overline{\Omega}}{\omega_R},\nonumber\\
g_j&=&\frac{a_{jj}}{a_0},\;\; g=\frac{a_{j,3-j}}{a_0}.
\label{g-wf}\end{eqnarray}
In the above, $E_R$ is the recoil energy and $a_0$ the background scattering length. Therefore, with
$\psi_j\equiv\psi_j(x,t)$, we obtain
\begin{eqnarray}
i \frac{\partial \psi_{j}}{\partial t} & =& \left[-\frac{\partial^2}{\partial x^2} + V(x) +  \Omega_j \right] \psi_{j}
- i b \frac{\partial \psi_{3-j}}{\partial x} + \nonumber \\
&& (g_j |\psi_{j}|^2+ {g} |\psi_{3-j}|^2) \psi_j, \;\;\;(j=1,2)
.\label{SO-GPE}
\end{eqnarray}
From (\ref{g-wf}) and (\ref{norm0}), the total number of atoms can be written
as
\begin{equation}
{N}= \frac{\omega_R}{2\omega_{\perp}k_La_0}\sum_{j=1}^2\int dx|\psi_j|^2=
\frac{\omega_R}{2\omega_{\perp}k_La_0}(N_1 + N_2),
\end{equation}
where $N_j$ represents the reduced fraction number of atoms in the component $j$.

A BEC system with a spin-orbit coupling as shown by the above formalism, when loaded in deep optical lattice can be
described in the tight-binding model with mean field approximation, by the following system for discrete nonlinear
Schr\"odinger  equation with SOC  (SOC-DNLS)~\cite{SA}:
\begin{equation}
\begin{split}
i\frac{d u_n}{dt}=&\,-\Gamma(u_{n+1}+u_{n-1})+ i\frac{\chi}{2}(v_{n+1}-v_{n-1})\\
&\,+\Omega u_n+ (\gamma_1 |u_n|^2+ \gamma  |v_n|^2) u_n, \\
i \frac{d v_n}{dt}=&\, -\Gamma (v_{n+1}+v_{n-1}) + i\frac{\chi}{2} (u_{n+1}-u_{n-1})\\
&\,- \Omega v_n + (\gamma  |u_n|^2+ \gamma_2 |v_n|^2) v_n,
\end{split}
\label{eq1}
\end{equation}
where
\begin{eqnarray}
\Gamma &\equiv& \Gamma_{n,n+1}=\int w^*(x-n)\frac{\partial^2}{\partial x^2}w(x-n-1) dx,\nonumber\\
\gamma &=& {\rm g} \int |w(x-n)|^4 dx,\;
 \gamma_i = {\rm g}_i \int |w(x-n)|^4 dx ,\\
\chi&\equiv& \chi_{n,n+1}= 2 b \int w^*(x-n)\frac{\partial}{\partial x} w(x-n-1) dx.\nonumber
\end{eqnarray}
Notice that, in the above system, we have  two conserved quantities:  the total number of atoms
\begin{equation}
N= \sum_n (|u_n|^2 + |v_n|^2),
\end{equation}
and the Hamiltonian
\begin{eqnarray}
&&H=\sum_n \left\{ - \Gamma(u_n^* u_{n+1} + v_n^* v_{n+1}) + i\frac{\chi}{2} u_n^{*}(v_{n+1} - v_{n-1})+ \right. \nonumber
\\&& \left.\frac{1}{4}(\gamma_1|u_n|^4+\gamma_2|v_n|^4)+\frac{\gamma}{2}|u_n|^2 |v_n|^2+\frac{\Omega_0}{2}(|u_n|^2-|v_n|^2)\right\} \nonumber \\ && +\,\, c.c.
\end{eqnarray}
where c.c. denotes the complex-conjugate of the expression in the curly bracket.

Next, in order to achieve a tunable SOC, we assume that the Zeeman field is
periodically varying in time, as
\begin{equation}
\Omega=\Omega(t)=\Omega_0 + \Omega_1 \cos(\omega t),
\label{Omega}\end{equation}
where $\Omega_0$ is the fixed constant part of the field and $\Omega_1$ the amplitude of the part
modulated with frequency $\omega$.
In view of this time-dependence of the Zeeman field, given by Eq.~(\ref{Omega}),
it is convenient to express the coupled system (\ref{eq1}) by an effective time averaged system,
which can be implemented by the following transformation:
\begin{equation}
u_n=U_n e^{-i\beta (t)}, \ v_n = V_n e^{+i\beta(t)},
\label{transform}
\end{equation}
where
\begin{equation}
\beta(t)= \Omega_1 \int_0^t \cos(\omega \tau) d\tau = \frac{\Omega_1}{\omega} \sin(\omega t).
\end{equation}
Once the transformation (\ref{transform}) is made, the coupled Eq.(\ref{eq1}) can be rewritten,
such that the explicit time dependence is removed from the Zeeman field (remaining only the constant
term $\Omega_0$), being transferred to the constant $\chi$, which has to be replaced by $\chi \exp(2i\beta(t))$.
Next, we perform the time averaging of Eq.(\ref{eq1}), over the period ($T=2\pi/\omega$) of the rapid oscillation,
by using that
\begin{equation}
\frac{1}{2\pi}\int_0^{2\pi}d(\omega t) \exp{\left(\frac{2i\Omega_1}{\omega}\sin(\omega t)\right)} = J_0\left(\frac{2\Omega_1}{\omega}\right),
\label{average}\end{equation}
where $J_0(\alpha)$ is the zero-order Bessel function in the variable  $\alpha$.
The above averaging procedure applied to Eq.~(\ref{eq1}), leads to the following coupled system:
{\small \begin{eqnarray}
i\frac{d U_n}{dt}&=&\,-\Gamma(U_{n+1}+U_{n-1})+ i\frac{\chi J_0(\alpha)}{2}(V_{n+1}-V_{n-1})\nonumber\\
&+&\Omega_0 U_n + (\gamma_1 |U_n|^2+ \gamma  |V_n|^2) U_n,\label{avsys} \\
i \frac{d V_n}{dt}&=&\, -\Gamma (V_{n+1}+V_{n-1}) + i\frac{\chi J_0(\alpha) }{2} (U_{n+1}-U_{n-1})\nonumber\\
&-&\Omega_0 V_n + (\gamma  |U_n|^2+ \gamma_2 |V_n|^2) V_n .\nonumber
\end{eqnarray}}
Quite remarkably, we see that the time averaged system given in Eq.~(\ref{avsys}) coincides with Eq.~(\ref{eq1}) under the
following replacement:
\begin{equation}
\Omega \to \Omega_0, \;\;\;\; \chi \to \chi_{eff} \equiv  \chi J_0(\alpha),\;\;\;\; \alpha\equiv \frac{2\Omega_1}{\omega}.
\label{eq8}
\end{equation}
Strictly speaking, these averaged equations are valid only in the strong modulation limit, e.g., when $\Omega_1$ and
$\omega$ are both very large with their ratio being finite. However, we shall see later that their
validity extends in a wide range away from this limit.

In the next two sections we study the spectral properties of the SOC system by diagonalizing the eigenvalue problem obtained
from the discrete coupled Schr\"odinger equation (\ref{avsys}) when we consider stationary solutions of the form:
\begin{equation}
U_n (t) = e^{-{\rm i}\mu t} U_n,\;\;\;\;  V_n (t) = e^{-{\rm i}\mu t} V_n,
\end{equation}
where $\mu$ is the chemical potential, related to the energy $E$ and to the total number of atoms $N$ by the relation $\mu=\frac{\partial E}{\partial N}$.

\section{Spectral properties of SOC tuned linear system}
\begin{figure}[t]
\begin{center}
\includegraphics[width=7cm]{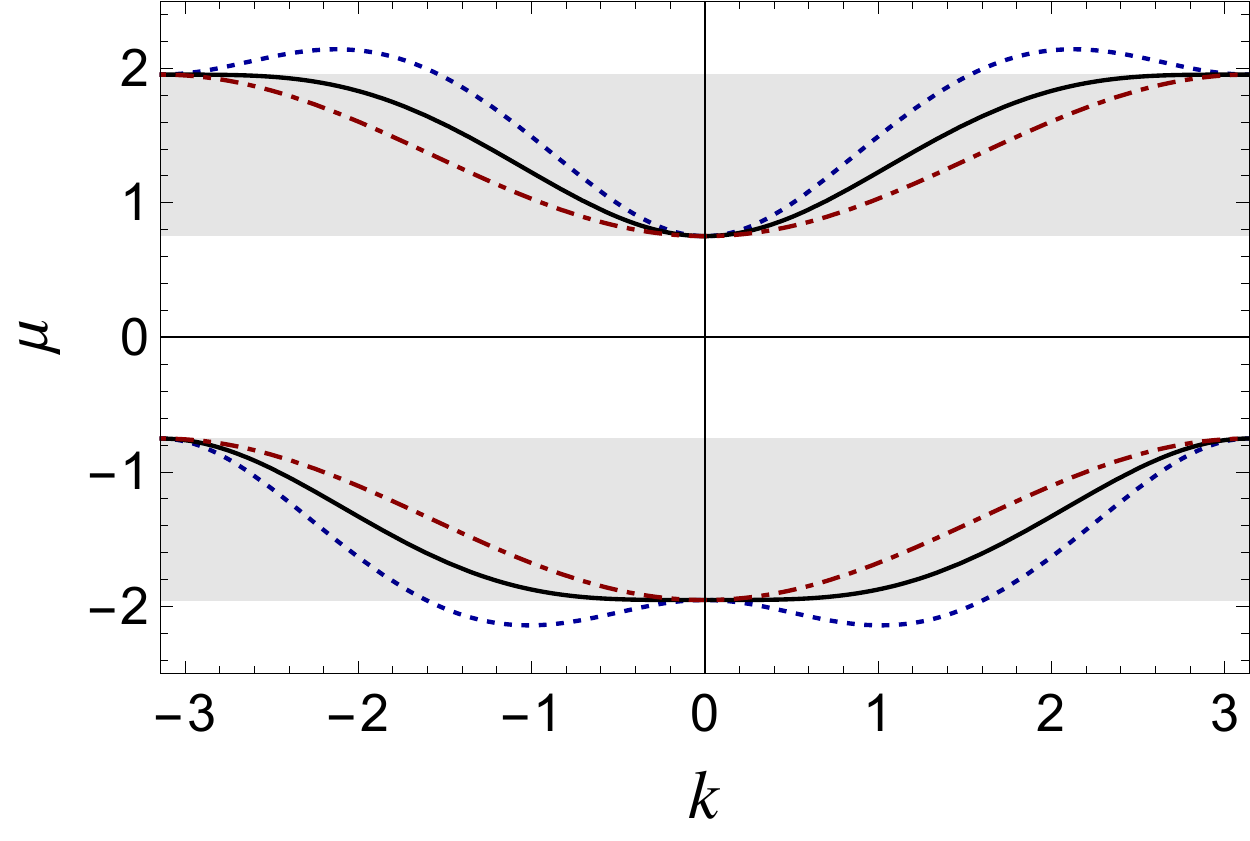}
\end{center}\vspace{-0.5cm}
\caption{Dispersion curves of the SOC-DNLS system for the linear case ($\gamma_1=\gamma_2=\gamma=0$), with three different
modulational parameters:  $\alpha=0.4$ (dotted-blue curves), $\alpha=1.33734$ (continuous curves), $\alpha=2.5$ (dot-dashed-red
curves).  Other  parameters are fixed as $\Gamma =0.3$, $\Omega_0=1.352$, $\chi=1.5$. Shadowed regions indicate the linear
bandwidths for values of $\alpha \ge \alpha_0^+$, where $\alpha_0^+\approx 1.33734$ is  the critical value in Eq.~(\ref{alfac}),
at which the local extremal point  coalesce.
}
\label{fig1}
\end{figure}

In the absence of any interaction, e.g. for $\gamma_1=\gamma_2=\gamma=0$,  the averaged system given
by Eq.~(\ref{avsys}) becomes exactly solvable and the dispersion relation  can be given analytically. Indeed,
from Eq.~(\ref{eq8}) we have  that  the linear dispersion relations of the modulated system simply follow from
the ones of  the unmodulated system given in~\cite{SA,BGPMHM}, as:
\begin{equation}
\mu(k,\alpha)_\pm = - 2 \Gamma \cos (k) \pm \sqrt{\Omega_0^2 + \left[\chi J_0(\alpha)\right]^2 \sin^2(k)},
\label{disp-rel}
\end{equation}
with  $k$, the crystal momentum,  varying in the first Brillouin zone $k\in[-\pi,\pi]$.
The minus and plus signs refer to the lower and upper parts of the dispersion curves (bands) in the reciprocal space, respectively.
 In Fig.~\ref{fig1} we depict the linear dispersion curves obtained from Eq.\ref{disp-rel} for three different values of the tuning parameter $\alpha$.
Starting from  Eq.~(\ref{avsys}), the behavior of the linear spectrum as a function of the tuning parameter $\alpha$, can be further
investigated. In this respect, notice from Eq. (\ref{disp-rel}) that the dispersion curves have two degenerate extremal points at
positions
\begin{equation}
k_{s}(\alpha)= \cos^{-1}\left[- s \frac{2 \Gamma}{\chi J_0(\alpha)} \sqrt{\frac{\Omega_0^2 + \left(\chi J_0(\alpha)\right)^2}
{4\Gamma^2 + \left(\chi J_0(\alpha)\right)^2}}\right]
\label{ks}\end{equation}
with $s=\pm 1$.

As the chemical potential for $k_s$ is given by
{\small
\begin{equation} \mu_s\equiv
\mu(k_s,\alpha)=
\frac{s\sqrt{\left[\chi^2 J_0^2(\alpha)+4\Gamma^2\right]\left[\chi^2 J_0^2(\alpha)+\Omega_0^2\right]}}
{\chi J_0(\alpha)},
\label{muks}\end{equation}}
the solutions at the extremal points are obtained from
 {\small
\begin{equation}
\frac{d\mu_s}{d\alpha}=
\frac{s}{\chi J_0^2(\alpha)}
\frac{J_1(\alpha)\left[\chi^4 J_0^4(\alpha)-4\Gamma^2\Omega_0^2\right]}
{\sqrt{\left[\chi^2 J_0^2(\alpha)+4\Gamma^2\right]\left[\chi^2 J_0^2(\alpha)+\Omega_0^2\right]}}=0,
\label{demudealfa}
\end{equation}}
which are at $\alpha_i=\alpha_i^\pm$ ($i=0,1,2,...$), given by
\begin{equation}
J_0(\alpha_i^\pm) = \pm \frac {\sqrt{2 \Gamma \Omega_0}}{\chi}
\label{alfac}
\end{equation}
and, at $\alpha=\eta_n$ ($n=0,1, 2,...$), for
\begin{equation}
J_1(\eta_n)=0.
\label{J1eta}
\end{equation}

\begin{figure}[t]
\begin{center}
\includegraphics[width=8.5cm,height=7cm]{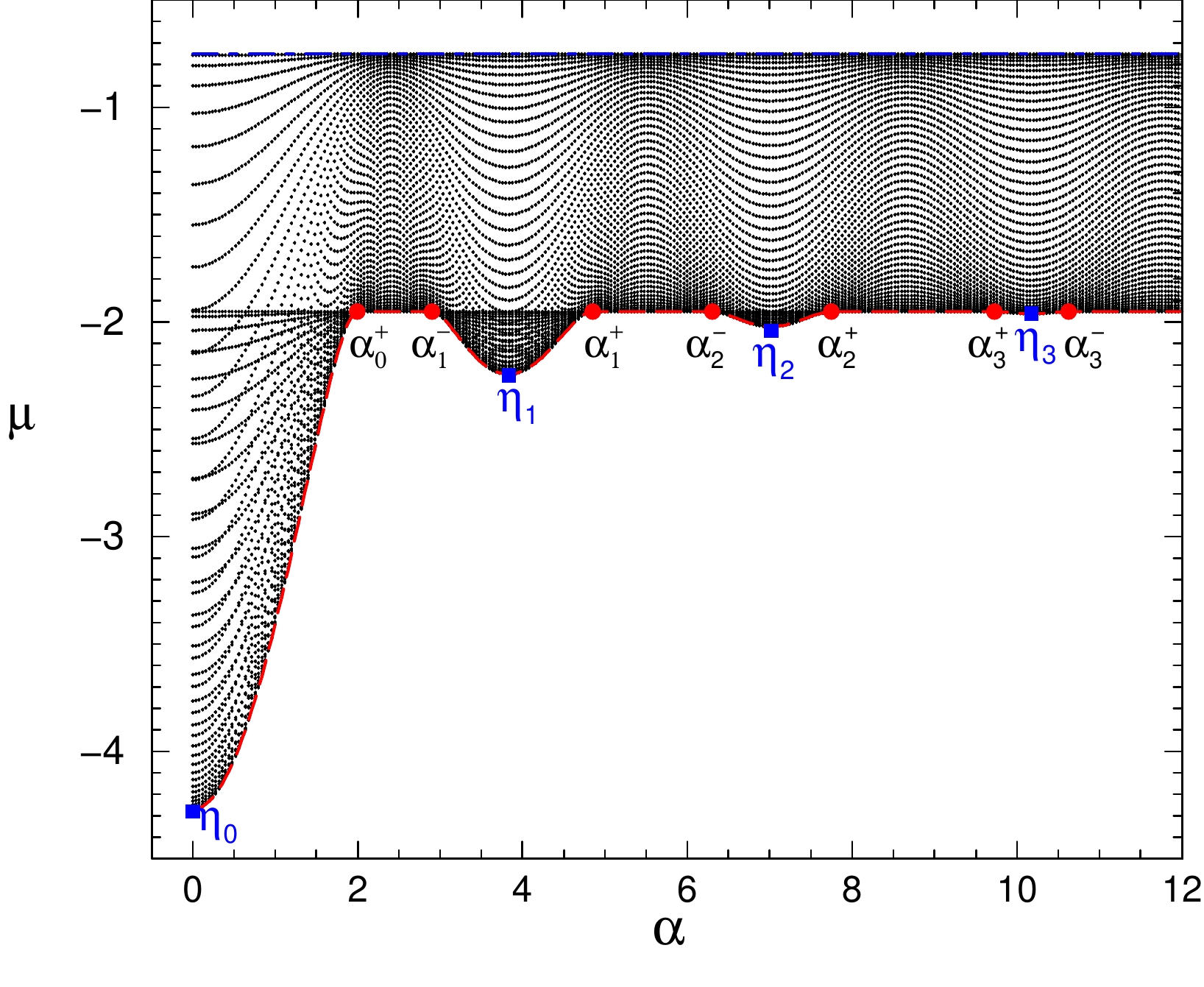}
\end{center}\vspace{-0.5cm}
\caption{
Lower-band spectrum of chemical potential as a function of $\alpha$ for a SOC-DNLS chain of 99 sites in the linear limit
$\gamma_1=\gamma_2=\gamma=0$, for $\chi=4.0$, with other parameters fixed as in Fig.~\ref{fig1}. 
The dashed-red curve displays the ground-state dependence, while the top dot-dashed-blue line corresponds to the
lower limit of the inter-band gap. For the chosen parameters we can identify one half lobe started at the origin (with
minimum at $\eta_0$, going till $\alpha_0^+$) and other three lobes (with minima at $\eta_1$, $\eta_2$, and $\eta_3$),
where the last one is hardly visible in the given plot scale.
Red-bullets and blue-squares correspond to values of $\alpha$ given by Eqs.~(\ref{alfac}) and (\ref{J1eta}), with
$\alpha_0^+=1.9978$, $\alpha_1^-=2.9023$, $ \alpha_1^+=4.8504$, $\alpha_2^-=6.3065$, $\alpha_2^+=7.7503$,
$\alpha_3^-=9.7298$, and $\alpha_3^+=10.6238$.
}
\label{fig2}
\end{figure}
A typical dependence of the linear spectrum
as a function of the tuning parameter $\alpha$  is  shown in Fig.~\ref{fig2}. In this figure, it is shown only part of the
spectrum corresponding to the lower band, since the part corresponding to the upper  band can be obtained from specular
reflection with respect to $\mu=0$ axis. Notice that different curves  correspond to different values of  $k$ and the spectrum
for a given $\alpha$ covers the first band in the  whole Brillouin zone $k\in [-\pi, \pi]$.

From Fig.~\ref{fig2} one can directly verify that the conditions given above are satisfied at the zeros of $J_1(\alpha=\eta_n)$,
given by (\ref{J1eta}), and for the possible solutions $\alpha_i^\pm$ of $J_0(\alpha)$, given by (\ref{alfac}). It is also easy to
check that for each  $\eta_i,\; i=1,2,3,...$, there exist  satellite solutions  $\alpha_i^-$,   $\alpha_i^+$ of Eq.(\ref{alfac}) lying
immediately before and after of $\eta_i$  and  equidistant from it, e.g. $\eta_i=(\alpha_i^{+}+\alpha_i^{-})/2$, while for the
point $\eta_0=0$
there exists  only the upper satellite $\alpha_0^+$ .
Thus, for all $\alpha \in R^+$ (notice that the dispersion relation is symmetric in $\alpha$),
the sequence $\alpha^*$ of all extremal points resulting from the above equations can be put in increasing order as follows
\begin{equation}
\alpha^*\equiv\{0, \alpha_0^+, \alpha_1^-, \eta_1, \alpha_1^+, \alpha_2^-, \eta_2, \alpha_2^+, ...  ,\}
\end{equation}
and thus the dependence on $\alpha$  of the extremal $\mu_{-}$ curve  can be separately  investigated for the sequence
of non overlapping intervals
\begin{eqnarray}
&& I_{\eta_0}= \left [0, \alpha_0^+ \right ], \,\,\,
I_{\eta_i}= \left [\eta_i -\Delta_i, \eta_i + \Delta_i \right ], \,\, i=1,2,... \nonumber \\
&& I_{\alpha_i}=]\alpha_i^+, \alpha_{i+1}^-[, \,\, i=0,1,2,...
\label{intervals}
\end{eqnarray}
with $\Delta_i=(\alpha_i^+ - \alpha_i^-)/2$.
{
One can prove that the chemical potential assume a constant value  $\mu_{-}=-2\Gamma-\Omega_0$ at the satellite points $\alpha_i^\pm$ and
inside all the intervals $I_{\alpha_i} \, i=0,1,2,..$.
This directly follows from Eq. (\ref{disp-rel}) and from  the fact that inside the intervals $I_{\alpha_i}$ the quasi momentum  $k_s(\alpha)$
becomes complex so that the only physical acceptable solutions for $\mu_-$ are the ones independent on $\alpha$, e.g. the ones for
which $k= 0, \pm \pi$   giving  $\mu=-2\Gamma\pm\Omega_0$ and $\mu=2\Gamma\pm\Omega_0$, respectively. Notice that, while the values
$-2\Gamma-\Omega_0$ and $2\Gamma+\Omega_0$ correspond, respectively, to the ground and to the highest excited states of the chemical
potential in the regions $I_{\alpha}$, the other two constants $-2\Gamma+\Omega_0$ and $2\Gamma-\Omega_0$ are delimiting the lower and
upper borders of the gap absolute. From this it follows that the lower and upper  extremal  curves are flat  for all
$\alpha \in I_{\alpha_i}$.
It is  worth to note that in terms of the dispersion curves in the reciprocal space, the critical values $\alpha_i^\pm$ also correspond to the
values of $\alpha$ for which the two minima (maxima)
$k_{-1}$  ($k_{1}$)  of  the lower (upper) band coalesce into a single minimum (maximum),
at $k=0$ ($k=\pi$).
This is pictorially illustrated in Fig.~\ref{fig1} where the linear dispersion curves are depicted    for different values of the tuning parameter $\alpha$.
\begin{figure}[t]
\centerline{
\includegraphics[width=4.2cm,height=3.8cm]{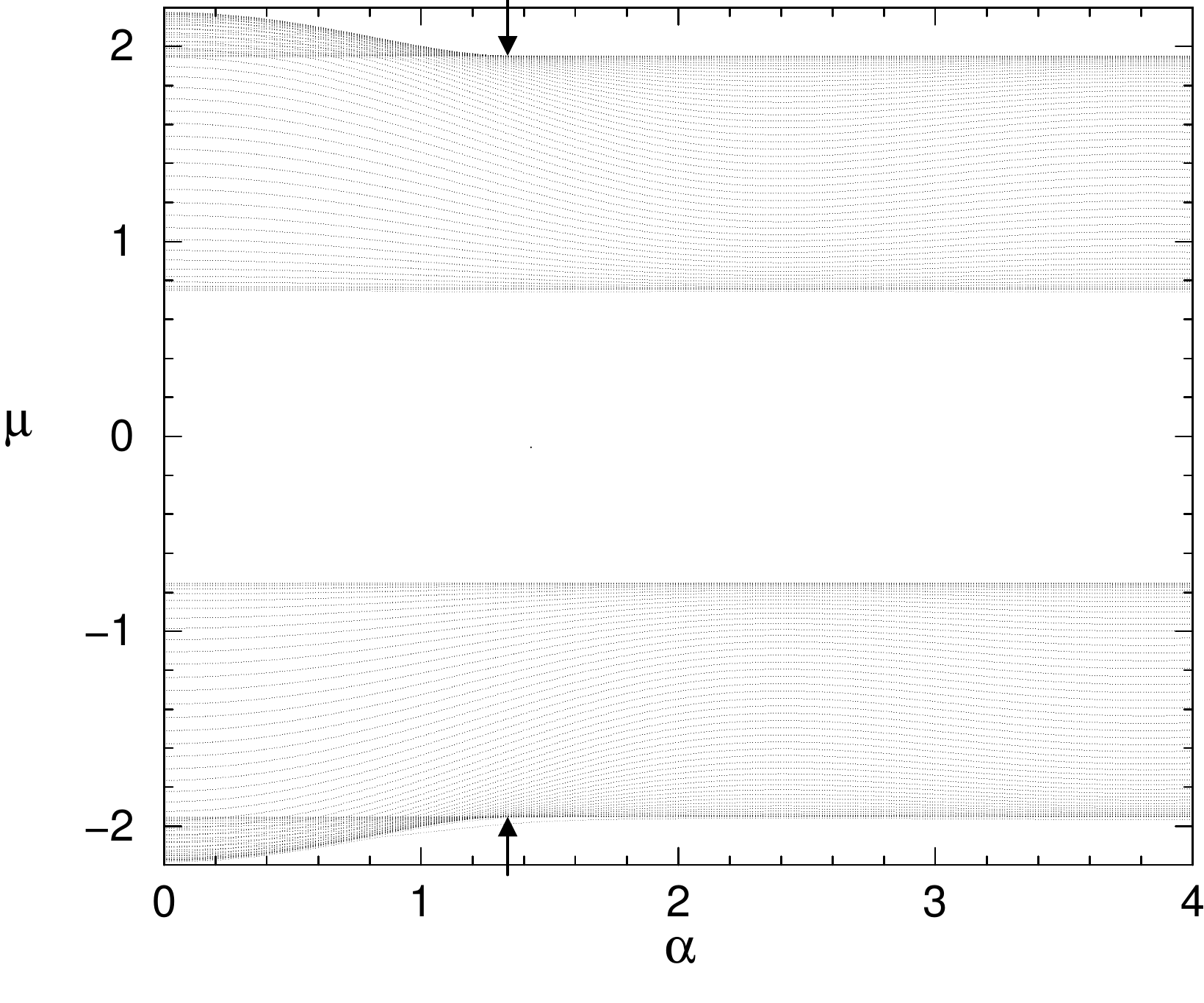}
\includegraphics[width=4.2cm,height=3.8cm]{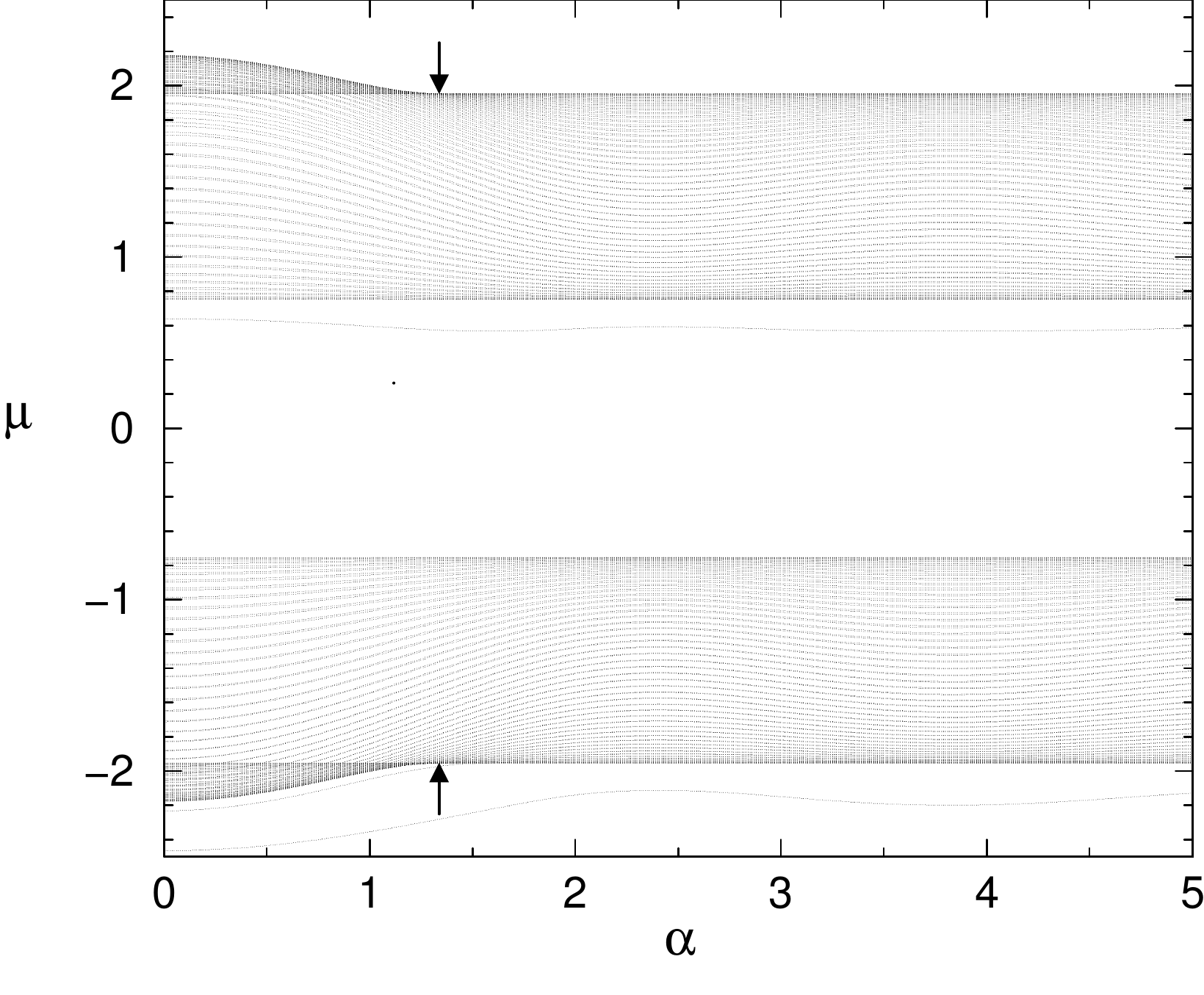}
}
\centerline{
\includegraphics[width=4.3cm,height=3.5cm]{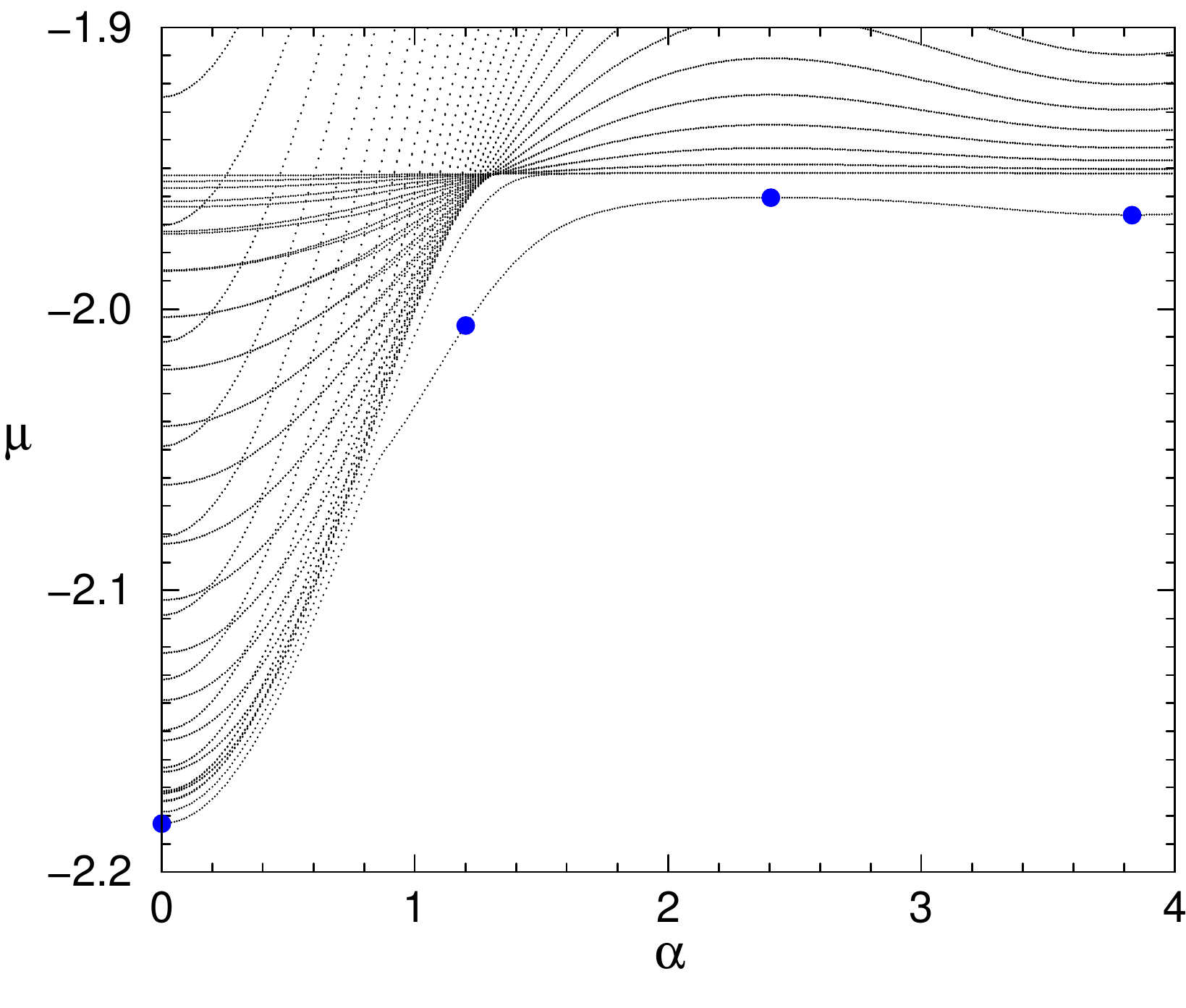}
\includegraphics[width=4.3cm,,height=3.5cm]{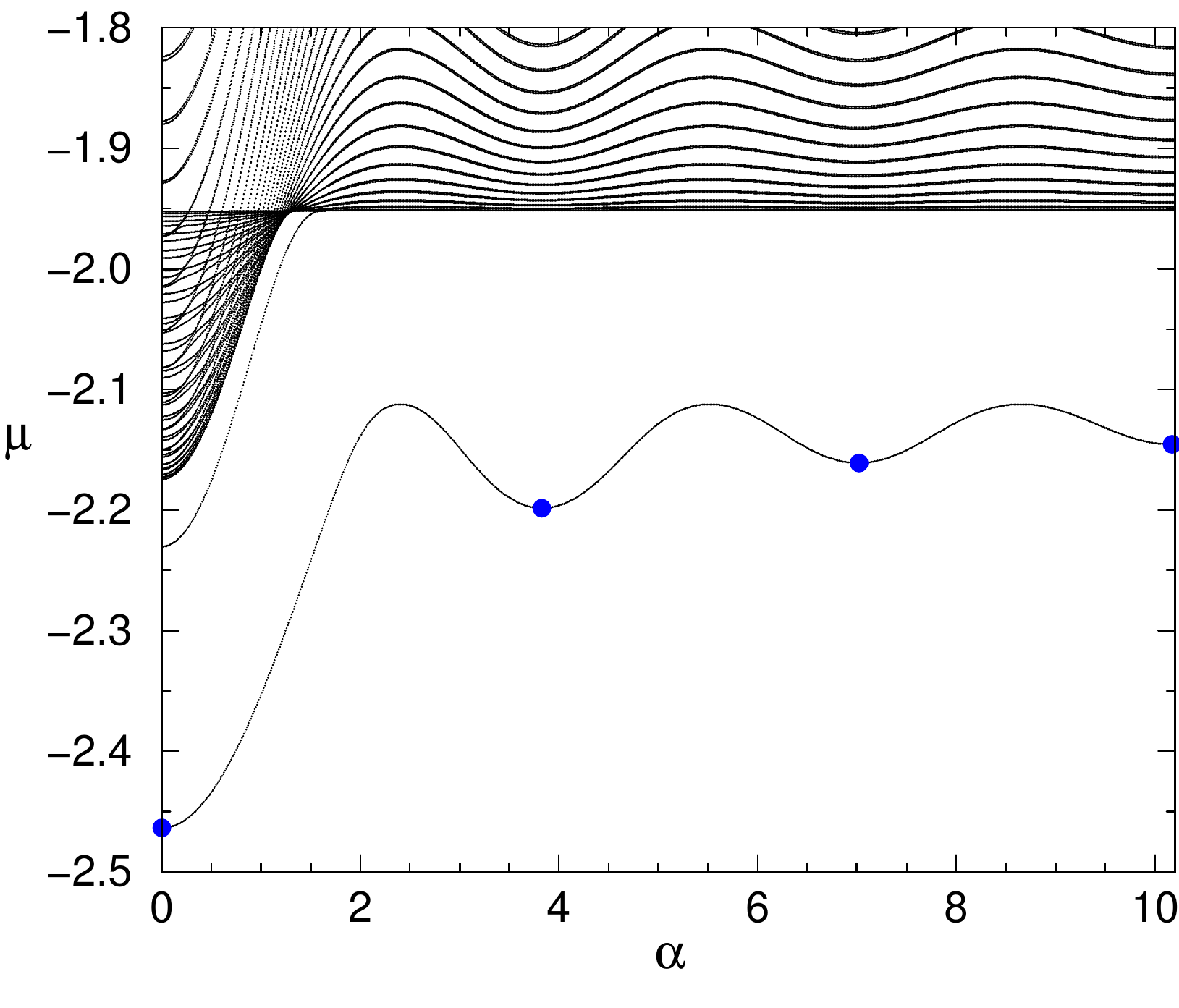}
}
\caption{
Energy spectrum  as a function of $\alpha$, within a chain with 99 sites, for the averaged SOC-DNLS system, with
$\gamma=-0.2$ (left panels) and $\gamma=-0.8$ (right panels).
In both cases, the full spectra are in the top panels, with the corresponding lower part in the bottom panels. 
The arrows in the top panels indicate the critical point $\alpha_0^+=1.33734$ at which the linear bandwidths becomes uniform
(see also Fig.~\ref{fig1} and Eq.(\ref{alfac}).
The other parameters are fixed as $\Gamma=0.3$, $\Omega_0=1.352$ and $\chi=1.5$. Notice that for these parameters only
the half lobe in the origin exists.
The blue bullets on top of the ground-state curves in bottom panels are related to the wave-functions shown
in Figs.~\ref{fig5} (right-panel) and~\ref{fig6} (left-panel).}
\label{fig3}
\end{figure}

On the other hand, in the  $I_{\eta_i}$ intervals, the dependence of the chemical potential on $\alpha$ gives continuous local  extremal
curves,  referred in the following  as ``lobes", which are symmetric  around their minimum at $\alpha=\eta_i$. The amplitude of the lobes
decrease as $\alpha$  is increased, the  absolute minimum being attained at  $\alpha=0$ where an half-lobe is observed (notice that
due to  the parity of $\mu$ on  $\alpha$ we can restrict only to non negative values of $\alpha$, meaning that lobe around
$\alpha=\eta_0=0$ becomes an half-lobe).

Also note  that the lobe profiles tangentially intersect the horizontal line $-2\Gamma-\Omega_0$  at the borders of the $I_{\eta_i}$ intervals.
From this it follows that the ground state curve and its derivative are both continuous functions of $\alpha$. These properties  can be directly
checked  by plotting  the curves $\mu_{-}(k_s,\alpha)$ in the interval $I_{\eta_i}$, with $s=(-1)^{i+1}$ for the i-th lobes,  $i=0,1,2,...$.

Thus, from the above analysis we conclude  that the ground state of the linear system is a continuous piecewise function of $\alpha$ which
consists  of a finite number of equally-spaced lobes at $\alpha=\eta_i$ (half lobe at $\alpha=0$) joined by the constant line
$\mu_{-}=-2\Gamma-\Omega_0$ inside the  $I_{\alpha_i}$ intervals. It can be proved that,for fixed values of the parameters, the number of
lobes in the ground-state curve (e.g., excluding  the half-lobe  at the origin) is given by the maximal  integer, $i_{max}$, for which the
quasi-momentum  $k_s(\eta_{i_{max}})$, with $s=(-1)^{i_{max}+1}$,  is still real. Therefore, the sequence of intervals in Eq. (\ref{intervals}) is
finite, with the last $I_{{\alpha}_{i_{max}}}$ flat interval given by $]\alpha_{i_{max}}, \infty]$.

Similar results follow by symmetry arguments also for  the  highest excited extremal curve $\mu_+(\alpha^*)$. In this case
$\mu_{+}=2\Gamma+\Omega_0$ at satellite points, lobes have   maxima at $\eta_i$ and tangentially intersect the constant line
$2\Gamma+ \Omega_0$ of intervals $I_{\alpha_{i}}$. Since the lower and upper border of the inter-band gap are constant in $\alpha$,
we also have that the lower band linear spectrum is constrained inside the lower extremal (ground state) curve and the lower gap border
$-2\Gamma+\Omega_0$ (similarly, the  upper band spectrum lies   between the upper gap border $2\Gamma-\Omega_0$ and the
highest excited state extremal curve).
In the next section we shall see that some of the linear features  survive also in the presence of nonlinearity.
}
\begin{figure}[t]
\centerline{
\includegraphics[width=4.2cm,height=3.8cm]{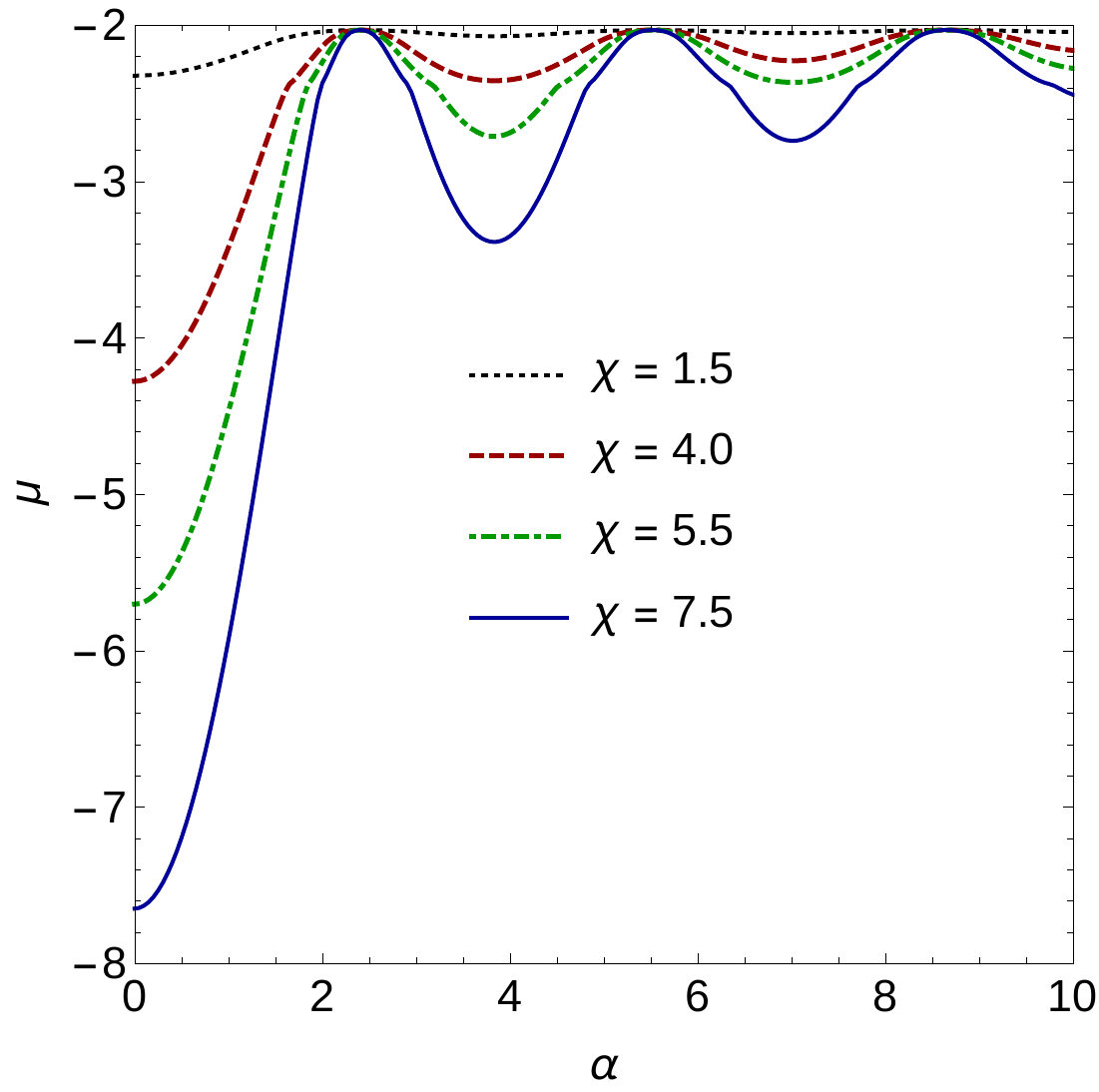}
\includegraphics[width=4.2cm,height=3.8cm]{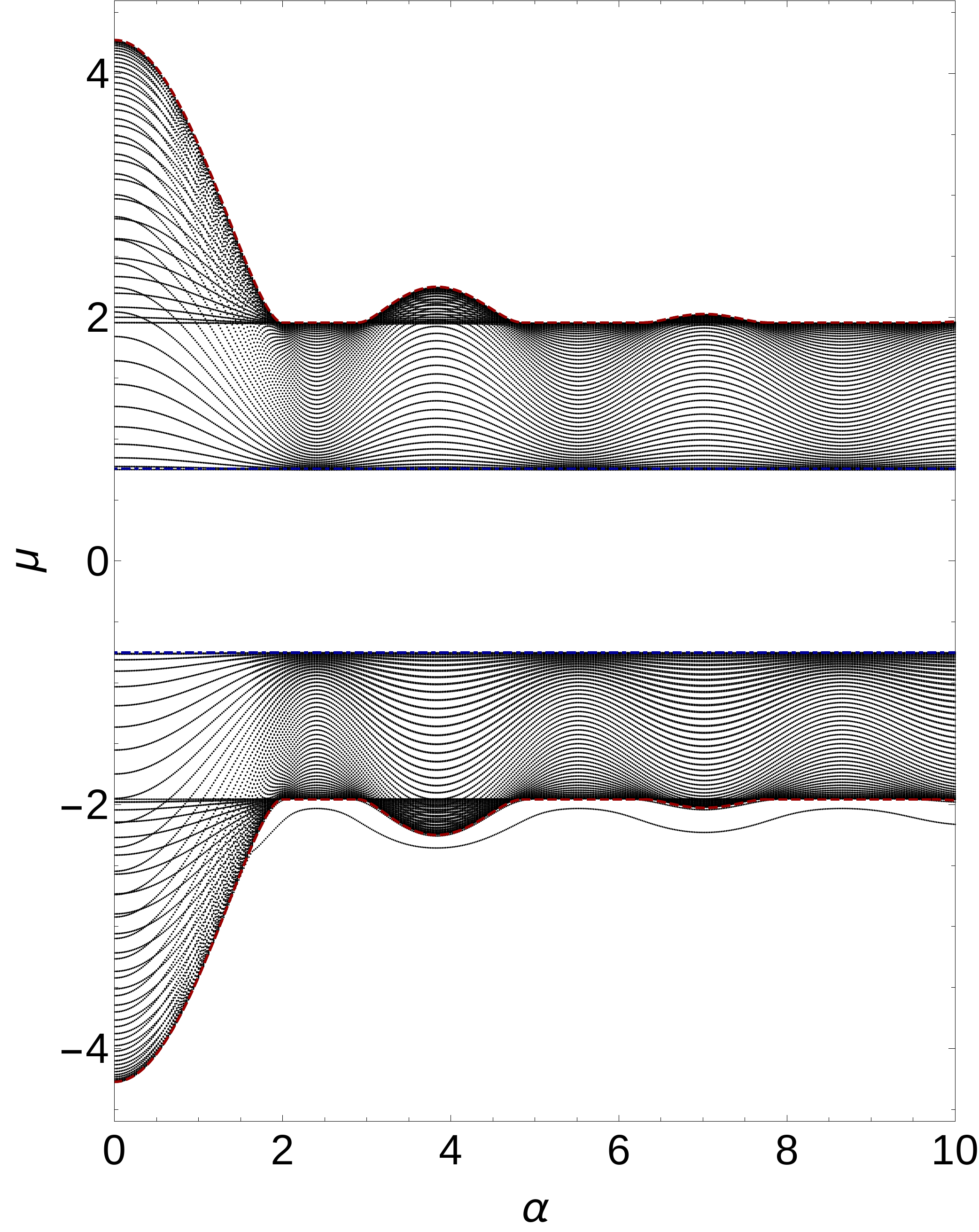}
}
\centerline{
\includegraphics[width=4.2cm,height=3.8cm]{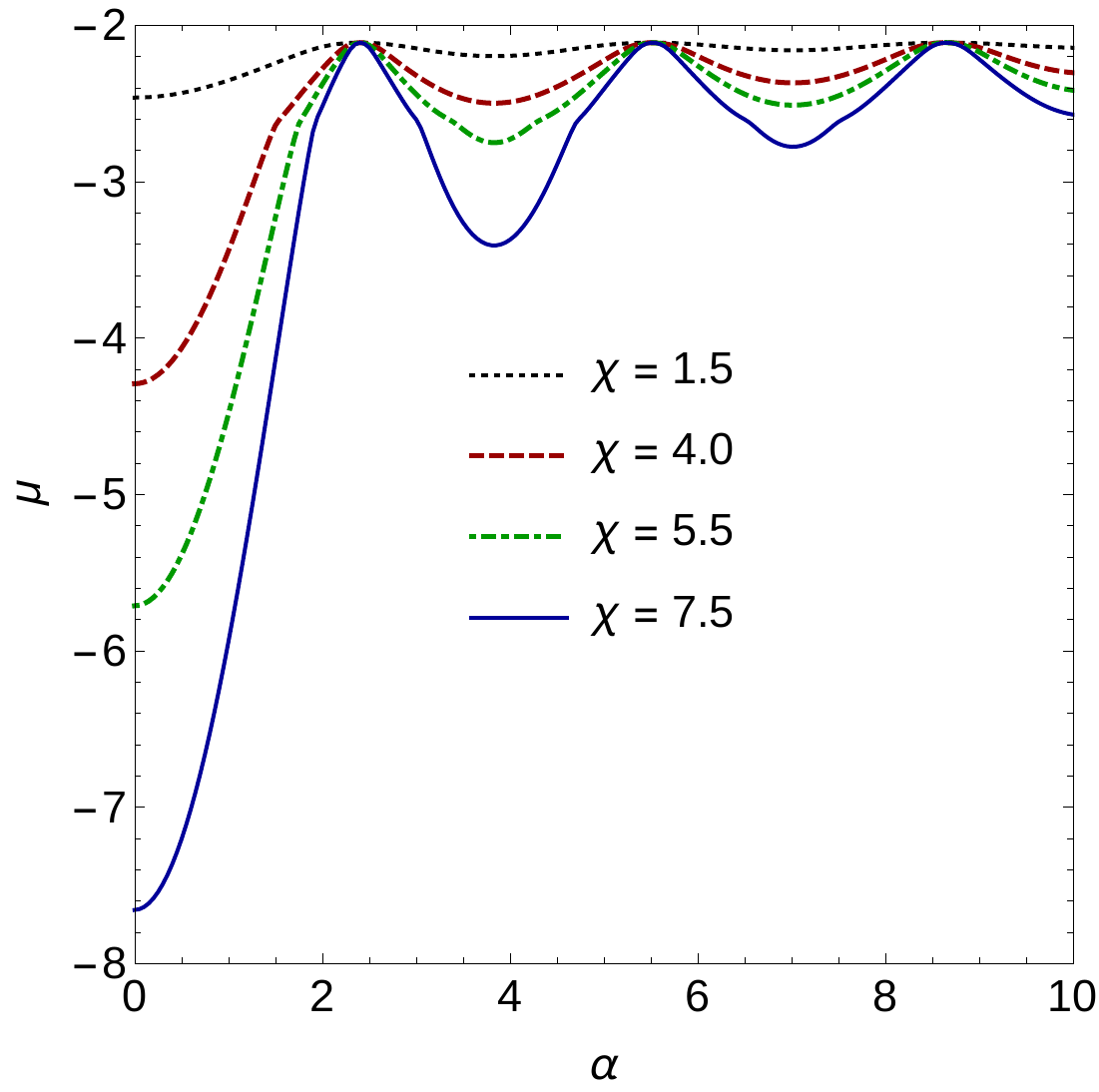}
\includegraphics[width=4.2cm,height=3.8cm]{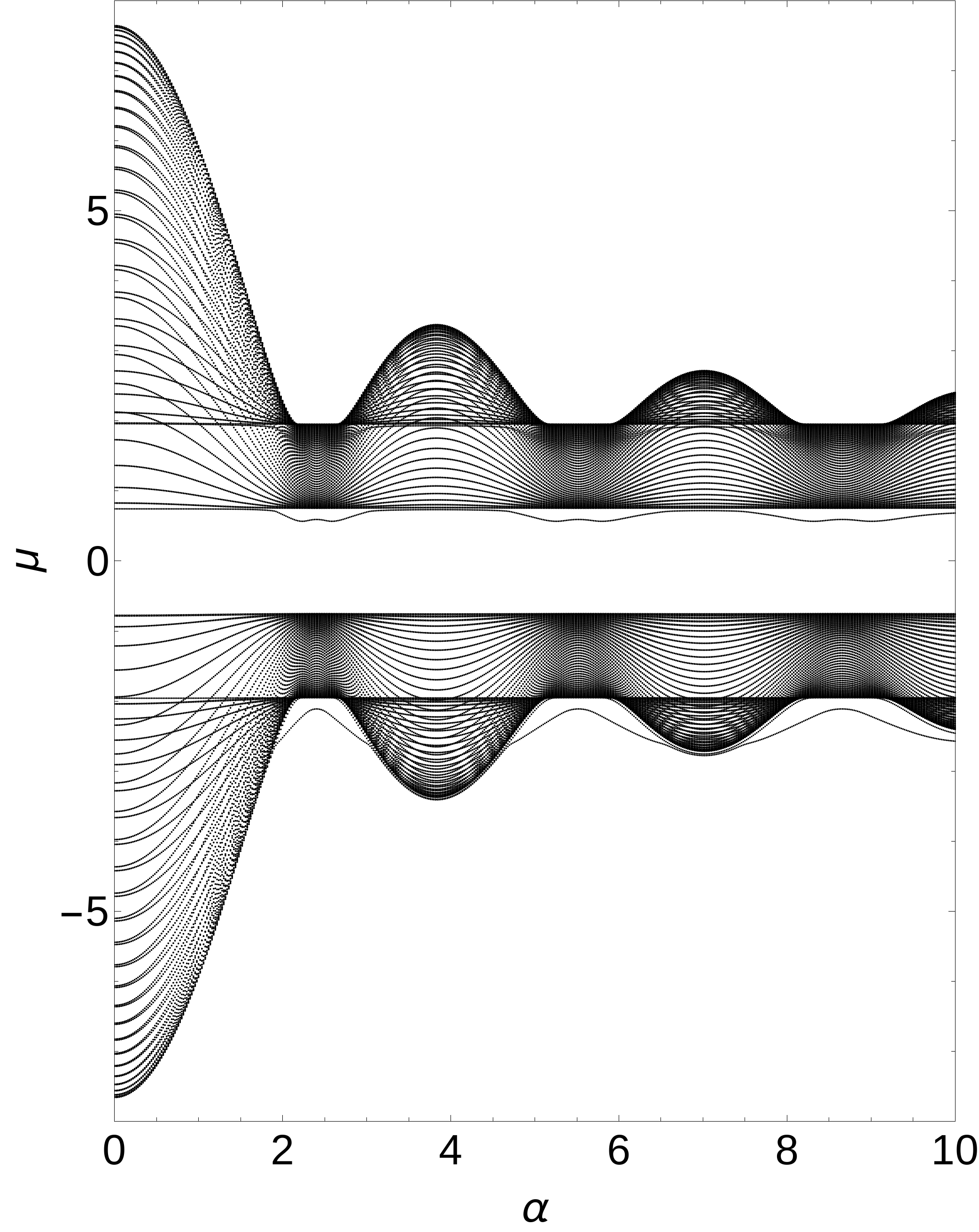}
}
\caption{Left panels: Ground-state chemical potentials as functions of the tuning parameter $\alpha=2\Omega_1/\omega$ for different values of $\chi$
(indicated inside the panels), with unequal, $\gamma_1=-0.4, \gamma_2=-0.6, \gamma=-0.1$ (top panel), and with all equal,
$\gamma_1=\gamma_2=\gamma=-0.8$  (bottom panel), nonlinearities. Other parameters are fixed as in Fig.~\ref{fig3}.
Right panels: Full spectrum vs $\alpha$ for $\chi=4.0$ (top) and  $\chi=7.5$ (bottom) with other
parameters fixed as in the corresponding left panel.
In the top right panel, the red lines refer to the ground-state and to the highest excited-state curves of the linear system,
respectively shifted by $0.05$  downward and upward  to avoid overlapping. Dashed-blue lines denote the constant bottom and
upper edges of the gap of the linear system. }
\label{fig4}
\end{figure}

\section{SOC tuned nonlinear spectrum}
Spectral properties of the nonlinear system have been  obtained from self-consistent exact diagonalization of the
averaged Hamiltonian system (\ref{avsys}). The numerical approach is described in more details in Ref.~\cite{ms05} for the  single component case, with
extension to multi-component case being straightforward. In the top panels of Fig.~\ref{fig3} we report  the chemical potential spectrum versus the tuning
parameter $\alpha$ as obtained for nonlinear cases with $\gamma=-0.2$ (left panels) and $\gamma=-0.8$ (right panels), considering all
equal attractive interactions with $\gamma_1=\gamma_2=\gamma$.
The top-left panel displays the behaviour of eigenvalues, which are quite similar to the linear case, with
the eigenvalues oscillating as functions of the tuning parameter, and with  amplitudes decreasing as $\alpha$ is increased.
One should notice that only one lobe at the origin appears for this set of parameters with $\chi=1.5$, in contrast with the linear
case shown in Fig.~\ref{fig2}, with $\chi=4.0$, where other lobes can be identified at the zeros of $J_1(\alpha)$, given by (\ref{J1eta}).
In the level of the oscillations, the position of the extremal points (maxima or minima) observed in the lower and upper
bands are in direct correspondence with the zeros of the Bessel function, $J_0(\alpha)$, and its first derivative, $J_1(\alpha)$. More explicitly, for the
case shown in the top-right panel, one can identify more clearly the corresponding ground state in the lower band, which is given in lower-right panel of
Fig.~\ref{fig3}). The observed minima are close to $\alpha=$ 0, 3.83, 7.02, 10.17 (zeros of $J_1(\alpha)$); with the maxima
close to $\alpha=$ 2.405, 5.52, 8.65 (zeros of $J_0(\alpha)$).

We should also observe that the upward (downward) rearrangement of the levels, giving rise to the half lobe of the nonlinear spectrum when the tuning parameter
is varied in the region $0 < \alpha \leq \alpha_0^+$, where $\alpha_0^+ = 1.33734$ is practically  the same  value expected for the  linear spectrum. Notice, however,
that the nonlinearity introduces  localized states in the band-gaps (this occurring at first order in the perturbation while  effects on band levels are  typically of higher
orders). Except for this, the qualitative  behavior of the spectral oscillations  (lobes) in the presence of nonlinearity can be qualitatively understood from the analysis
performed in the previous section. In particular, note from  Eq.~(\ref{disp-rel}) and from the crossover of the linear bands across the critical point in Eq. (\ref{alfac})
that, for $\alpha<\alpha_0^+$,  there are points of the  spectrum lying outside the shadowed region of Fig.~\ref{fig1}. These  points correspond to the upper and lower
band lobes observed in the top panels of Fig.~\ref{fig3}. However, for $\alpha\ge\alpha_0^+$, all points lie inside the shadowed region corresponding to the flat curves
shown in Fig.~\ref{fig3}, in full agreement with the analysis of the linear system, in spite of the presence of the nonlinearity (for the chosen parameters the linear
spectrum has only the half lobe at the origin and  the flat  semi-infinite interval $]\alpha_0^+, \infty]$).

Spectral modulations induced by the Zeeman term in the presence of nonlinearity are also depicted in the bottom panels of  Fig.~\ref{fig3} for two different sets of
nonlinearity parameters. As shown, in these cases the main difference is the  appearance of isolated levels in the gaps. The spectral oscillations inside the bands
persist in presence of nonlinearity, with the ground-state curve oscillating in phase with curves of excited levels inside the lower band.

The effects of the nonlinearity on the ground-state energy  and on the full spectrum  are further investigated as functions of $\alpha$ in Fig.~\ref{fig4}.
In particular, in the left panels of this figure we show ground-state behaviors for different values of the spin-orbit parameter $\chi$, with  two different choices of
the nonlinear parameters corresponding to attractive interactions. In the upper-left panel we have all unequal interactions and, in the lower-left panel, all the
interaction parameters are the same.
The full spectra with respect to $\alpha$, for two specific values of  $\chi$, with all  other parameters as in the corresponding left panels
 are shown in the right panels of Fig.~\ref{fig4} . From this figure, we notice that the behaviors for ``all equal" and ``all unequal" interactions  are  qualitatively
 similar, this being particularly true if nonlinearities are not too large. Moreover,  the amplitude of the oscillations increase with the increasing of $\chi$, as
 verified for the ground state, which is a natural consequence of the $\chi$ dependence on the  rescaling (\ref{eq8}).
 In contrast with the linear case, the ground-state curves display points where the derivative changes abruptly; a phenomenon becoming more
 evident for larger values of $\chi$. For instance, see the case with $\chi=7.5$ at the bottom-left panel of Fig.~\ref{fig4}. These points are in correspondence
 with values of $\alpha$ where a localized level in the semi-infinite gap touches a  band lobe (say, the $i-$th lobe), with subsequent detachment at a
 point $\alpha$ symmetrically located with respect to $\eta_i$ (this being particularly visible for  the first lobe of the $\chi=7.5$ spectrum).
 At these points, a SOC induced change of symmetry properties occurs, similar to the one reported in Ref.\cite{SA}.
 The localization of the ground state changes rapidly at such points, passing from a well localized state inside the gap to a nonlinear stripe-like  extended state
 bordering the lobe band (see bottom right panel of Fig.~\ref{fig4}  and Fig.~\ref{fig9} below).

In conclusion, as far as the nonlinear spectrum is concerned, we can say that the main role of  the nonlinearity is to introduce localized states in the gap,
which display very interesting change of properties when they undergo collisions  with the band lobes. Remarkably, the structure of the extremal curves
(including the gap) of the linear band is well preserved also in the presence of intermediate (not too large) values of the nonlinearity (For instance,
compare the top right panel of Fig.~\ref{fig4} with Fig.~\ref{fig2}).

With respect to the localized  states in the band-gaps, they refer to discrete versions of gap-solitons of the continuous BEC mixtures in OLs.
Their existence is related to the modulational instability of linear Bloch states~\cite{KS01}, a well known phenomenon that we are not discussing here.
Existence and stability of SOC tunable discrete solitons  will be instead investigated in the next section by numerical methods.  In view of the qualitatively
similar results observed for different nonlinearity values, in the rest of this paper we refer only to attractive and all equal magnitude interactions.
\begin{figure}[t]
\centerline{
\includegraphics[scale=0.24,width=4.2cm,height=4.7cm]{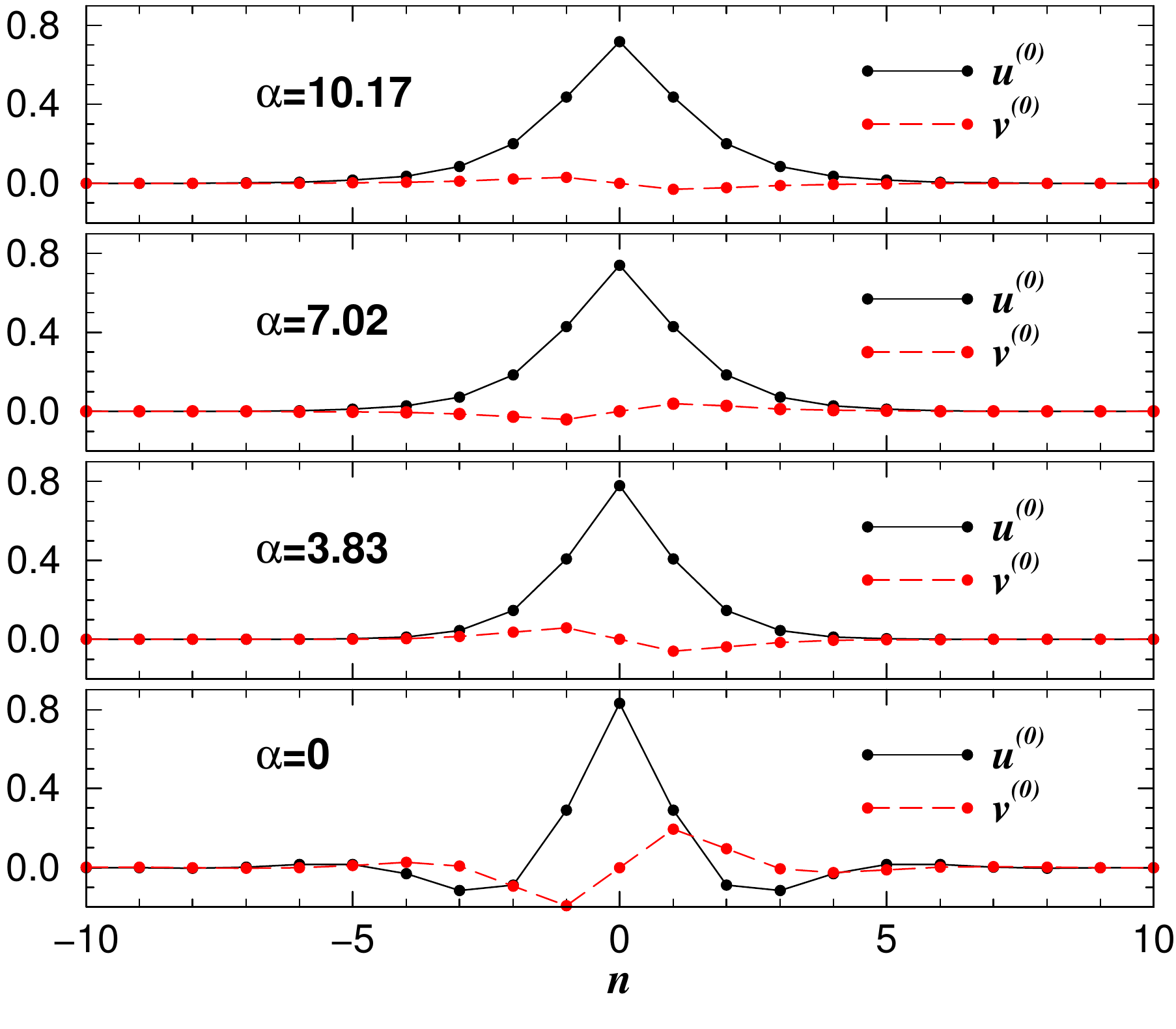}
\includegraphics[scale=0.24,width=4.2cm,height=4.7cm]{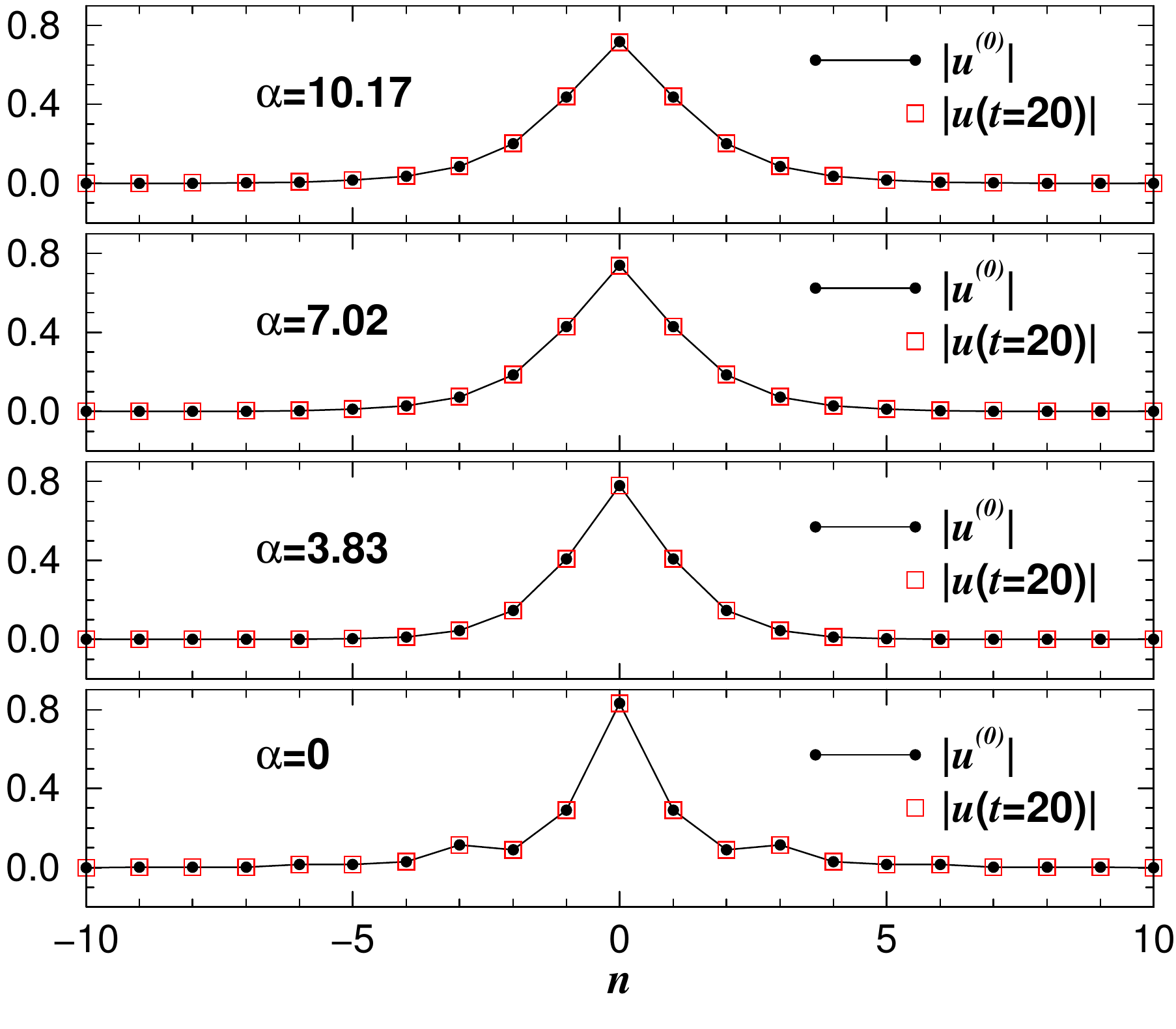}
}
\caption{
Ground-state wave functions related to the four minima depicted in the right-lower panel of
Fig.~\ref{fig3}, corresponding to the zeros of $J_1(\alpha)$ (given inside the panels).
In the left panels we have the $u$ (pure real) and $v$ (pure imaginary) components. In the right
panels, we indicate the time-stability of the results by considering just the module of the $u$
component, for $t=0$ and $t=20$. The parameters are as in the
right frames of Fig.~\ref{fig3}, with $\Omega_1=100$, $\Omega_0=1.352$, $\gamma=-0.8$,
$\Gamma=0.3$ and $\chi=1.5$.
 The wave function is normalized as in Eq. (\ref{norm}), with the respective number fractions
 of the $u-$component given by $N_u=$ 0.9056,  0.9897, 0.9950, and 0.9967.}
\label{fig5}
\end{figure}

\begin{figure}[h]
\centerline{
\includegraphics[width=4.2cm,height=4.7cm]{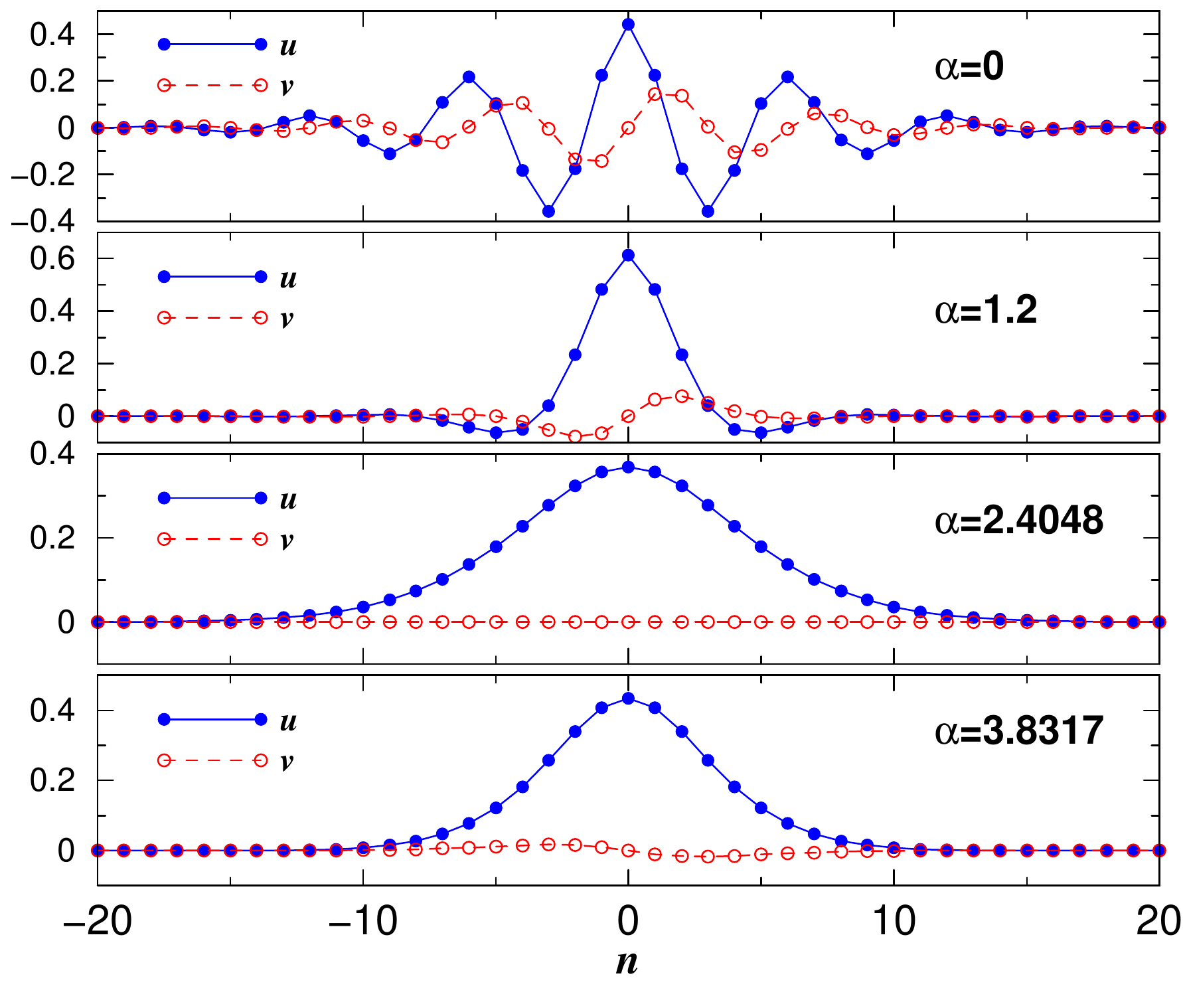}
\includegraphics[width=4.2cm,height=4.7cm]{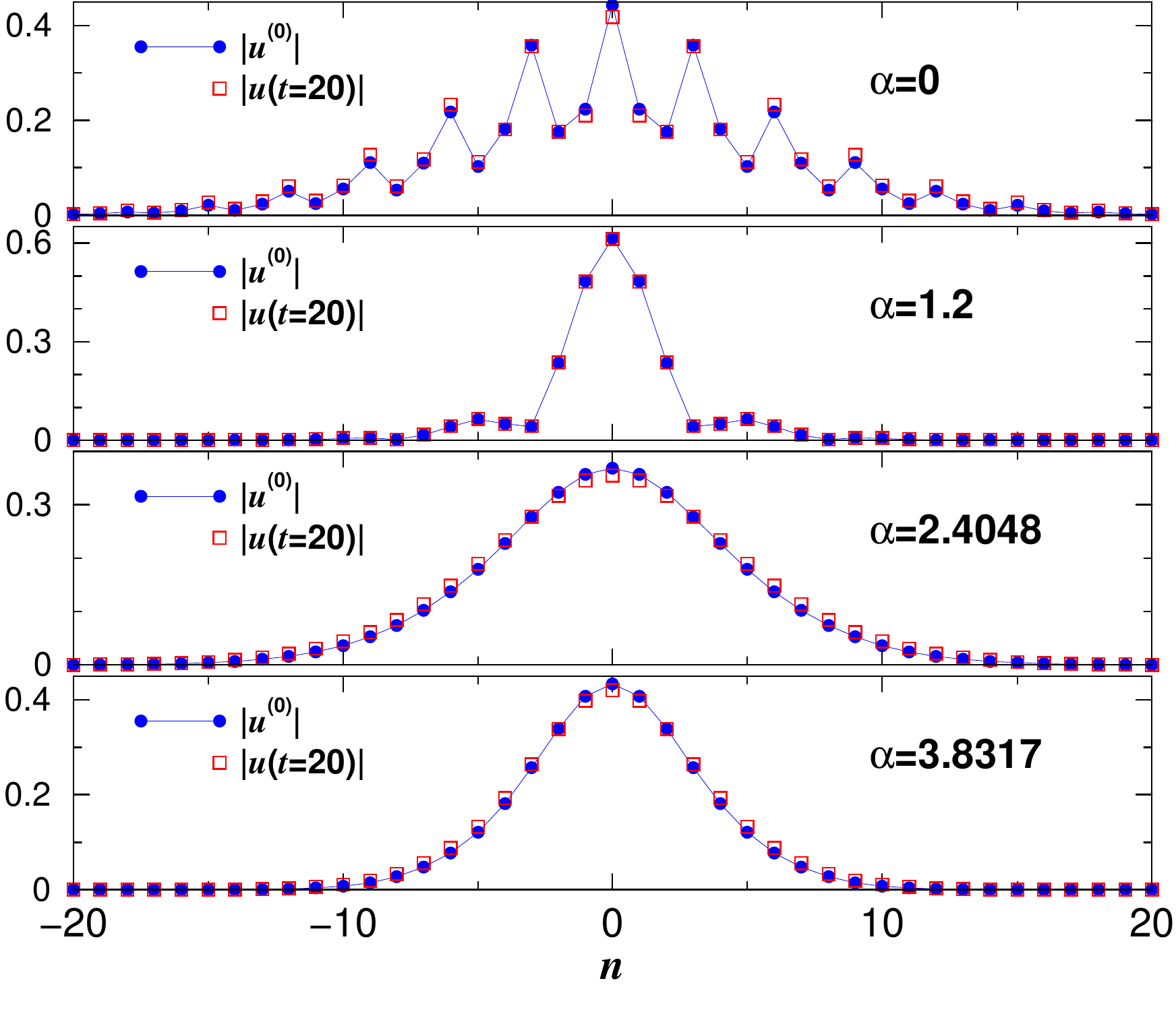}
}
\caption{Ground-state wave functions corresponding to the four dots depicted
in the left-lower panel of Fig.~\ref{fig3}, for the $\alpha$ values indicated inside the
frames. In the left panels we have the components $u$ (solid-blue) and $v$ (dashed-red).
In the right panels, the time stability is indicated by considering the module of $u$ for $t=0$
and $t=20$. The parameters  are as in the left frames of Fig.~\ref{fig3}, with $\Omega_1=100$,
$\Omega_0=1.352$, $\gamma=-0.2$, $\Gamma=0.3$ and $\chi=1.5$.}
\label{fig6}
\end{figure}

\section{SOC tuned DNLS solitons}
In this section we consider effects of the SOC tuning on the existence, stability and localization properties of  stationary solitonic ground
states and stripe solutions of both averaged and  original (e.g., with time modulated Zeeman term) systems. To this regards,  we
recourse to numerical methods which we briefly describe here. For the averaged system, besides the self consistent numerical
diagonalization to obtain spectral properties discussed in the previous section, we also consider the relaxation method based
on imaginary time evolution~\cite{suzuki-varga} with a 4th order Runge-Kutta (RK) method to obtain the ground-state wave functions,
with periodic boundary conditions. In the imaginary time evolution, and in all our numerical calculations,  the components
$u_n$ and $v_n$ of the eigenstates were normalized with respect to the total wave function,
\begin{equation}
\sum_n (|u_n|^2 + |v_n|^2)=1.
\label{norm}
\end{equation}
The results obtained with imaginary time propagation were  found in perfect agreement with the ones obtained by
self-consistent method and presented in Fig.~\ref{fig3} for the ground state.

Real time evolution is also performed with the same RK code, with time step up to 10$^{-4}$, and the same periodic
boundary conditions. During the real time evolution, the conservation of the total norm was always monitored to check the
accuracy.

In the left panel of Fig.~\ref{fig5} we depict  the stationary ground states of the averaged system in correspondence of the four local
minima $\alpha=\eta_i$ ($i = 0,1,2,3$) represented in the energy curve displayed in the bottom-right panel of Fig.~\ref{fig3}. As expected,
for attractive interactions these ground states are found to be stable under time integrations of the averaged equation Eq.~(\ref{avsys}),
as well as under  time evolutions of the full system, with $\Omega_1=100$ and $\omega$ fixed according to the given value of $\alpha$.
In Fig.~\ref{fig5}, the stability under time evolution is evidenced in the corresponding right panels, where we show results for the
absolute values of the $u-$component, considering $t=$0 and 20.
Notice, from the corresponding left panels, the change of internal phase of the wave function at different minima, with the tendency
to become more localized at small  values of the tuning parameter, expanding as $\alpha$ increases. It is worth to remark that
maxima of the oscillating part of the ground-state curves  are in correspondence to the Bessel function $J_0$ zeros. Therefore,
correspond to the  vanishing of the rescaled SOC parameter.  Ground-state profiles at these points are obviously less localized,
since their chemical potentials have minimal distance from the linear band. On the contrary, for  $\alpha=0$ we have $J_0(\alpha)=1$,
having the largest value of the SOC parameter. One could expect the state to be more localized at this point. The maximal localization,
however, is achieved somewhere between $\alpha=0$ and the first zero of $J_0$ as a result of the interplay between SOC and
nonlinearity. Similar behaviors are found also for a different (lower) value of the nonlinearity as one can see from the panels in
Fig.\ref{fig6}, corresponding to the ground-state curve shown in the  bottom left panel of Fig.(\ref{fig3}). As one can observe
from these Figs.~\ref{fig5} and \ref{fig6}, the maximal localization is also achieved at the intermediate value $\alpha \approx 1.25$.
\begin{figure}[t]
\includegraphics[scale=0.42]{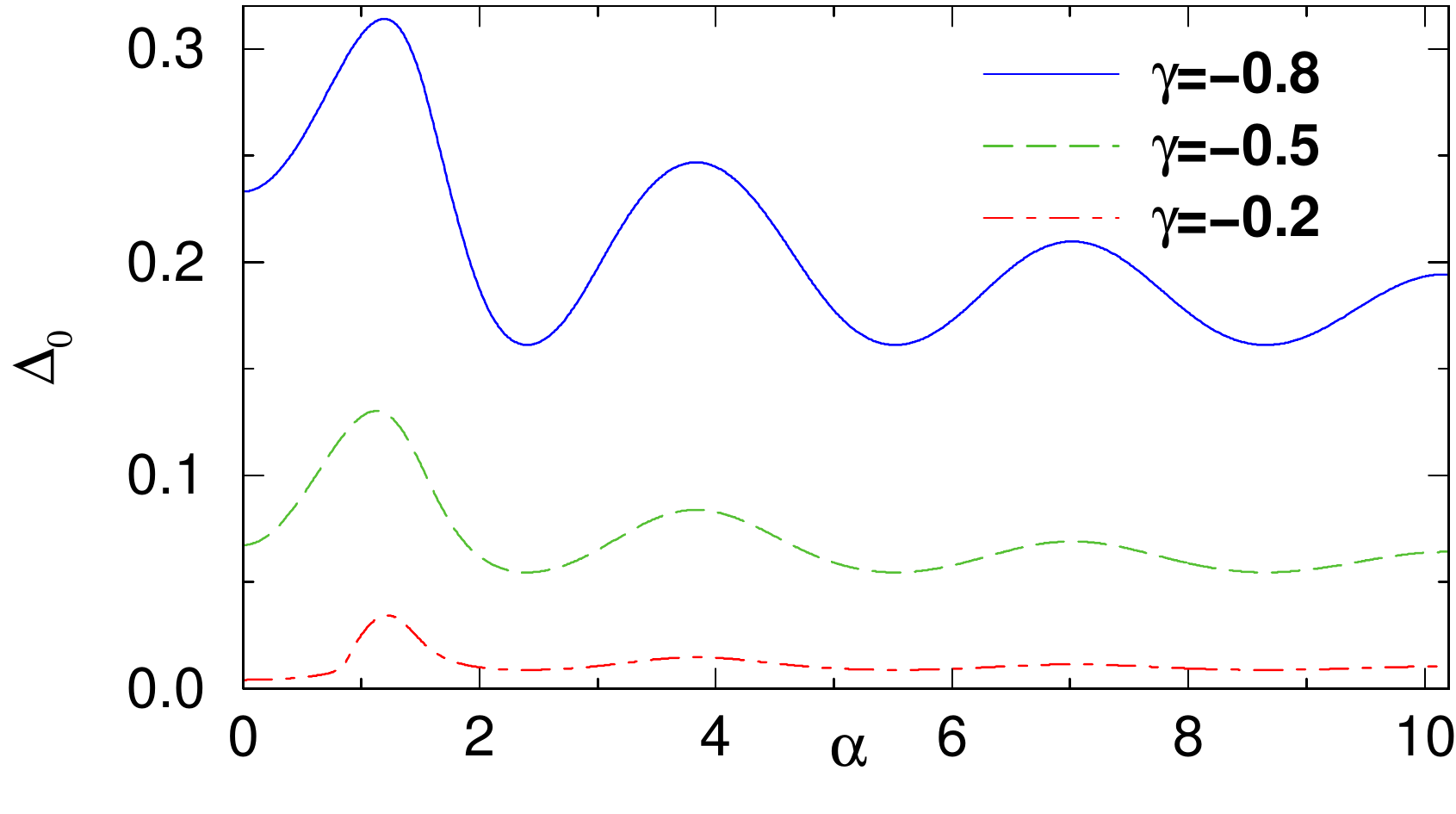}
\caption{
Ground-state gap energy, defined by $\Delta_0\equiv \mu(1)-\mu(0)$, as a function of the tuning parameter $\alpha$,
for different attractive interactions $\gamma$, as identified inside the frame.
The first peaks (where we have the maximum deviation) of the curves occur at $\alpha=1.22$ (for $\gamma=-0.8$),
$\alpha=1.13$ (for $\gamma=-0.5$) and $\alpha=1.23$ (for $\gamma=-0.2$).
The other parameters are fixed as in the bottom panels of  Fig.~\ref{fig3}.}
\label{fig7}
\end{figure}

In order to better quantify the influence of the SOC modulation on the ground-state localization, we have depicted in Fig.~\ref{fig7}
the behavior of the ground-state gap energy, $\Delta_0$, as a function of $\alpha$ for three different values of the interatomic
interaction parameter $\gamma$, where $\Delta_0$ is defined as the difference between the ground state and the first excited
state of the lower band.
We observe that, for $\gamma=-0.2$, the maximum gap is achieved for $\alpha \approx 1.23$, in correspondence to the intermediate value
between $\alpha=0$, where SOC parameter is maximum,  and $\alpha=\eta_1$,  where the corresponding ground-state curve has its first local minimum.

One should also observe that, as it is natural to expect for attractive interactions, the gap $\Delta_0$ increases as the interatomic
interactions increases,  but the relative weight of the peak at $\alpha\approx 1.3$ becomes more pronounced at small nonlinearities.
The peak is a consequence of the interplay of  SOC and the nonlinear interactions.
Since at the largest value of $\Delta_0$ the chemical potential of the ground state is more  detached from the linear band, it is clear
that at this value one expects the  maximal localization. For the chosen parameters, this  is achieved at $\alpha \approx 1.2$,
with a very small dependence on the interaction parameter $\gamma$, as one can see from Fig.~\ref{fig7}. A similar behavior
 is found  also for the excited localized wave functions inside the inter-band gap; however, we do not pursue the
analysis of these states here, because they appear to be unstable under time evolution.

\begin{figure}[t]
\centerline{
\includegraphics[scale=0.23]{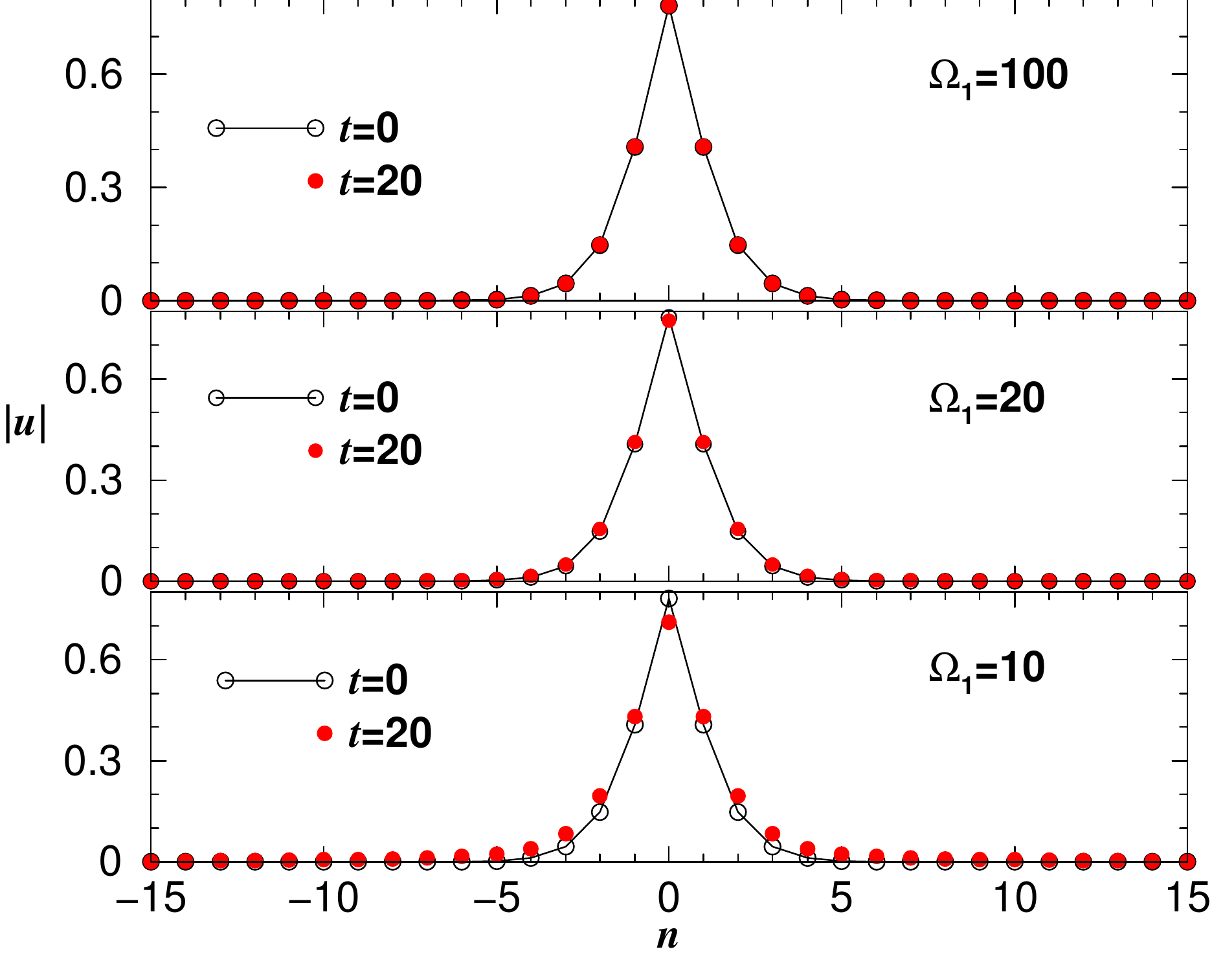}
\includegraphics[scale=0.23]{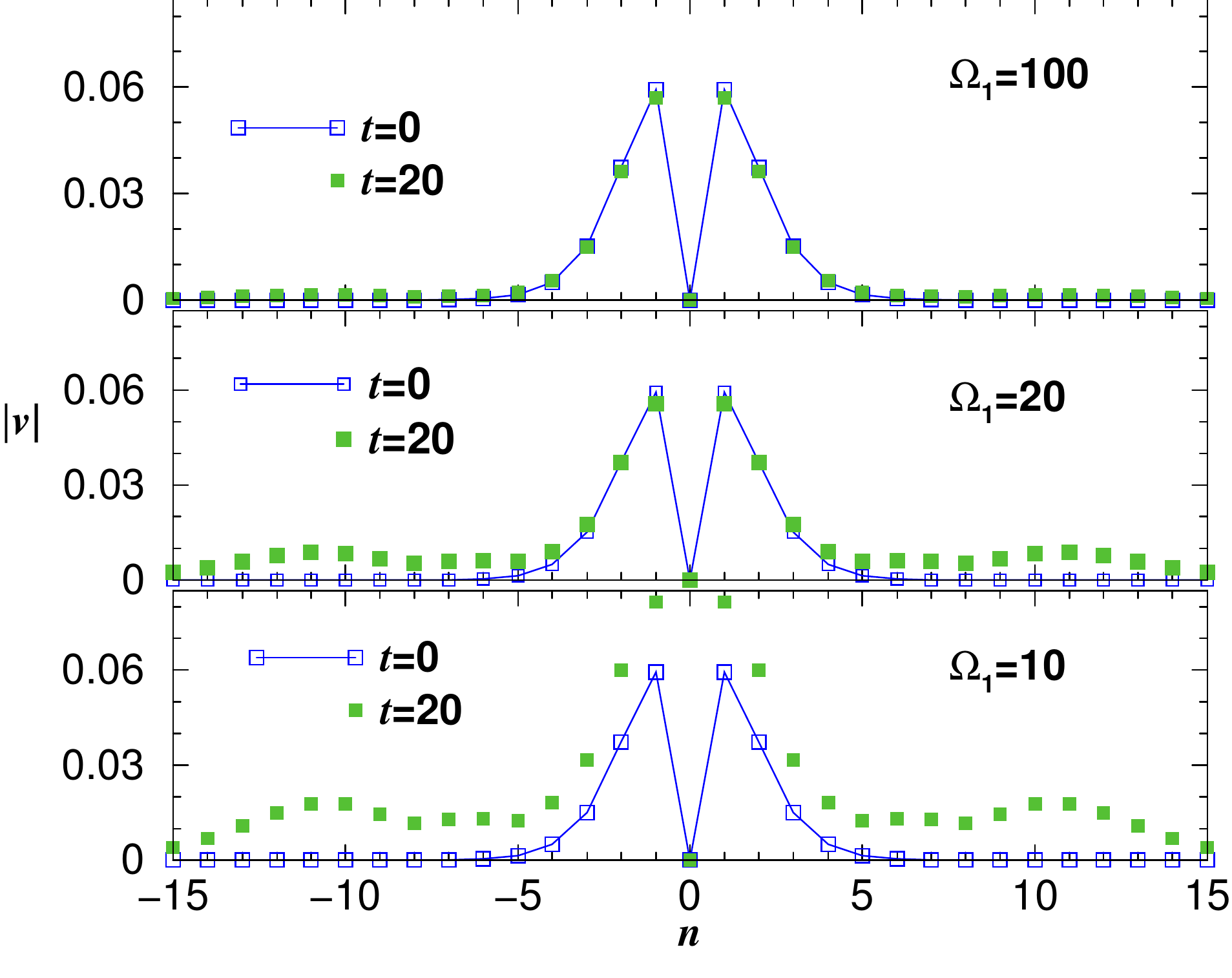}
}
\caption{
Full numerical simulations of $|u|$ (left panels) and $|v|$ (right panels) in the ground state,
for different amplitude oscillations $\Omega_1=$10, 20 and 100 (as shown explicitly), with nonlinearity fixed by
$\gamma=-0.8$. The corresponding chemical potential is $\mu=-2.20$, with the other parameters being
$\Omega_0=1.352$, $\Gamma = 0.3$, $\chi=1.5$ and $\alpha=$ 3.83.
In all the simulations, we start relaxation with $\chi_{eff}=\chi J_0(\alpha)=-0.604$ ($t=0$), performing
real-time evolution of Eqs.~(\ref{eq1}) with $\chi=1.5$.
}
\label{fig8}
\end{figure}
We have also investigated the range of validity of the averaged equations away from the strong
modulation limit, with results presented in Fig.~\ref{fig8}. In this respect, we consider, for a fixed value of $\alpha$, the original
time modulated system with different oscillation amplitudes $\Omega_1$, ranging from very large to relatively small values, with the
corresponding frequency $\omega$ fixed by the chosen $\alpha$. We  use the exact solution of the averaged system as
 initial condition to start the time propagation under Eq.~(\ref{eq1}). We found enough
illustrative to present the stability results for the absolute values of the ground-state components $u$ and $v$, by considering
three fixed values of $\Omega_1$ ($=$10, 20 and 100), with the time evolution being performed from $t=0$ till $t=20$.  As shown
from the lower panels of Fig.~\ref{fig8} (better visualized from the quite smaller values of the component $v$), the results for
$t=20$ start to deviate from original one when we have $\Omega_1=10$, increasing the discrepancy for smaller values of this
amplitude.
We can see from this figure that  in the strong modulation limit the eigenmodes of the averaged system are excellent solutions of
Eq. (\ref{eq1})  for $\Omega_1=100$, remaining good even largely below this value (some deviation in the $v$ component start
to appear around $\Omega_1=20$. From this we conclude  that, although from a strict mathematical point of view the averaged
theory is  valid for $\Omega_1, \omega \rightarrow \infty$, the  range of applicability of our results is quite large and is likely to
be within the present experimental feasibilities.

We remark that, besides the stationary ground states considered above, it is also possible to have nonlinear ground-state solutions
resembling stripe solutions of the linear system. In this case, stripes are linear superpositions of the degenerated ground states with
opposite quasi-momentum  in the lower branch of the dispersion curve (see Fig.~\ref{fig1}). These states can exist also in the presence
of nonlinearity, although not as exact linear combinations, as they have more complicated format. They can be constructed  as long
as quasi double degenerated minima in the dispersion curve survive in presence of nonlinearity (this is true for weak nonlinearities).
From numerical point of view, they can be constructed from exact stripes of linear system, continuing then by path following method
as the nonlinearity is increased.

\begin{figure}
\centerline{
\includegraphics[width=8cm]{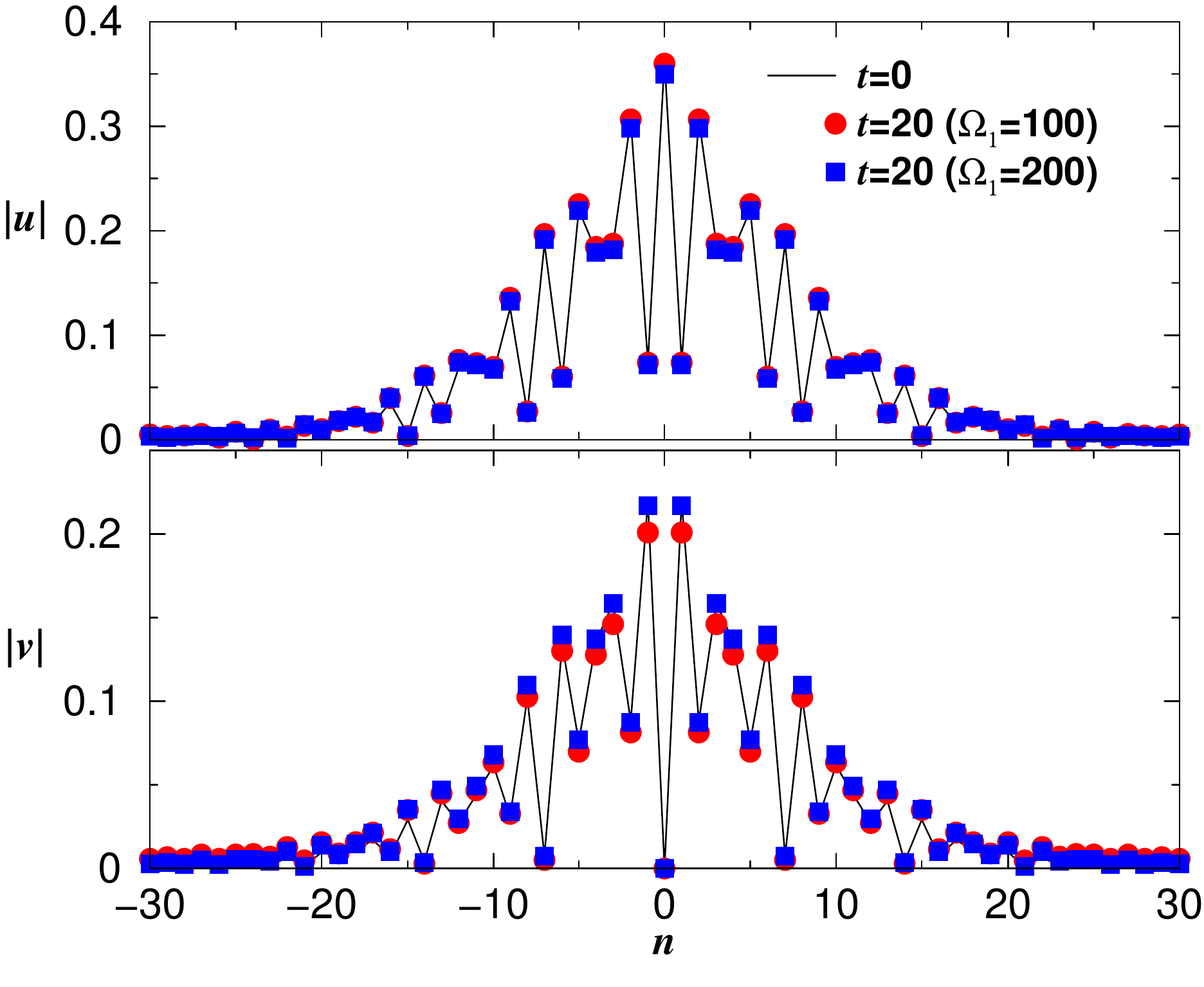}
}
\caption{
Ground-state wave-function components, $u$ (upper panel) and $v$ (lower panel),
for stripe-soliton solutions from full numerical simulations, for the case with $\chi=7.5$.
The corresponding chemical potential is $\mu=-3.41$.
In both the panels, with black-solid lines we have the case with $t=0$.
For $t=$20, we verify the stability of the results for $\Omega_1=$100 (red bullets) and
$\Omega_1=200$ (blue-squares).  Except for $\chi=7.5$, the other parameters are as
in Fig.~\ref{fig8}.
In the simulations, the relaxation is done with $\chi_{eff}=-3.021$, and perform
real-time evolution of Eqs.~(\ref{eq1}) with $\chi=7.5$.
}
\label{fig9}
\end{figure}

In Fig.~\ref{fig9}, we show results obtained for a stripe-like soliton, with $\chi=7.5$, and for two large
values for the amplitude, $\Omega_1=$100 and 200 consistent with the ratio, $\alpha=2\Omega_1/\omega=3.83$
as in Fig.~\ref{fig8} and at the position $\alpha=\eta_1$.  Notice that, for the above values the modulated linear
dispersion curve has two minima in the lower branch, such that one can directly check the results from the exact
modulated dispersion relation, which assures the existence of linear stripe solution for these parameter values.

From the above results we conclude that the modulation of the Zeeman term can be effectively used for the tuning of the
SOC parameter via a simple rescaling in Eq.~(\ref{eq8}); and, in turn, this permit to control the energy and the localization
properties of  the ground-state wave functions.

\section{Discussion and Conclusions}
Before our concluding remarks, we shall  briefly discuss a parameter design for possible experimental observation of
the above results. In this respect we refer to the SOC for the case of $ ^{87}$Rb atoms in the field of three laser beams
implemented in  a tripod scheme. The ground states from the $5S_{1/2}$ manifold are coupled via differently polarized
light, by chosing $|1\rangle=|F=2,m_F = -1\rangle $, $|2\rangle  = |F=2,m_F =+1\rangle $ and
$|3\rangle  = |F=1, m_F =0\rangle $~\cite{EOUFTO}. A deep
optical lattice can be  induced by additional two contra-propagating laser beams of strength of the order $\approx 10$
recoil energy. The number of atoms can be taken as  $N_0 \approx 3\cdot 10^3 $, with lattice wavelength
$\lambda_L = 1\mu$m, radial trapping frequency  $\omega_{\perp}\approx 10^3 $Hz  and $a_0 =100a_B$ (with $a_B$,
the Bohr radius), $\omega_R=2\cdot 10^5$Hz.
The strong modulation limit can be reached by considering a modulated Zeeman field of  normalized amplitude $>20$
and frequency of the modulation fixed by $\omega=2 \Omega_1/\alpha$. Under these circumstances, it should be possible
to check our results; and, in particular, the localized properties of the ground stated at specific values of the
modulation parameter discussed above.

In conclusion,  we have investigated the effect of a modulating Zeeman field on the energy spectrum and on the eigenstates
of binary BEC mixture in a deep OL and in the presence of SOC, by considering an exact self-consistent numerical diagonalization of the
averaged Hamiltonian. Stationary solitonic ground states and stripe modes are also investigated as functions of the modulating parameter,
both by exact diagonalizations and  by imaginary time evolution.
In particular, we derived proper averaged equations and showed  that the chemical potentials of solitonic states display oscillatory behaviors
as a function of the tuning parameter $\alpha$, whose amplitudes decrease as $\alpha$ is increased.
The dependence of the spectrum on the tuning parameter has been fully characterized for the linear SOC system. In this case, the
dispersion relations were exactly derived and the extremal curves (ground and highest excited states) of the linear system were shown
to be  continuous functions, together with their derivatives, consisting  of a finite number of band lobes joined by  constant lines.

The linear case for BEC with SOC can be experimentally realized, when the interactions are tuned to negligible quantities, by using
Feshbach resonance technics, i.e., by the variation of the external magnetic field near the resonant value.
As for the nonlinear spectrum, it is shown that the main role of the atomic interactions is to introduce localized states in the band-gaps,
which  undergo changes of properties as they collide  with the lobes. Remarkably, the structure of the extremal curves of the linear band
is well preserved also in the presence of nonlinearity (at least, when such nonlinearities are not too large). The ground-state stability
in the presence of a modulating field was demonstrated by real time evolutions of the original (non-averaged) system.

Finally, we remark that the control of the  localization properties of the ground state of a  BEC mixture in a deep optical lattice
by means of the SOC parameter could be very useful for applications involving soliton dynamics, including  nonlinear Bloch oscillations,
dynamical localization,  and interferometry. By following the present approach, indeed,  one could adjust the Zeeman field so to
achieve the maximal localization of a soliton ground state without changing the inter and intra-species interactions.

\section*{Acknowledgements}
M.S. acknowledges partial support from the Ministero dell'Istruzione, dell'Universit´a e della Ricerca through a
Programmi di Ricerca Scientifica di Rilevante Interesse Nazionale initiative under Grant No. 2010HXAW77-005;
FA acknowledges support from Grant No. EDW B14-096-0981 provided by IIUM(Malaysia) and from a senior visitor fellowship
from Conselho Nacional de Desenvolvimento Cient\'\i fico e Tecnol\'ogico (CNPq-Brasil). AG and LT also thank the Brazilian
agencies CNPq, Funda\c c\~ ao de Amparo \`a Pesquisa do Estado de S\~ao Paulo (FAPESP) and Coordena\c c\~ao de
Aperfei\c coamento de Pessoal de N\'\i vel Superior (CAPES) for partial support.

\end{document}